\documentclass[reprint,superscriptaddress,amsmath,amssymb,twocolumn]{revtex4-2}

\usepackage{amsmath}
\usepackage{mathtools}
\usepackage{amssymb}
\usepackage{graphicx}
\usepackage{bm}
\usepackage[colorlinks, linkcolor=blue, citecolor=blue, urlcolor=blue, breaklinks=true]{hyperref}
\usepackage{color,dsfont,multirow,booktabs}
\usepackage{braket}
\usepackage{tabularx}
\usepackage{lipsum,booktabs}
\usepackage{subfiles}
\usepackage{subcaption}
\usepackage{caption}
\usepackage[dvipsnames]{xcolor}
\usepackage{soul}
\usepackage{ragged2e}
\usepackage{orcidlink}
\usepackage[T1]{fontenc}
\usepackage[normalem]{ulem}%

\newcommand{\commentold}[1]{}
\DeclareMathSymbol{:}{\mathpunct}{operators}{"3A}

\newcommand{\corr}[1]{#1}
\newcommand{\del}[1]{\sout{#1}}

\begin{document}
\date{\today}

\title{Hybridized-Mode Parametric Amplifier in Kinetic-Inductance Circuits}

\author{Danial Davoudi}
 \author{Abdul Mohamed}
\author{Shabir Barzanjeh}
 \email{shabir.barzanjeh@ucalgary.ca}
\affiliation{%
Department of Physics and Astronomy, University of Calgary, Calgary, AB T2N 1N4, Canada}%

\begin{abstract}
Parametric amplification is essential for quantum measurement, enabling the amplification of weak microwave signals with close to the minimum added noise permitted by quantum mechanics. While Josephson-junction--based amplifiers have become standard in superconducting quantum circuits, their magnetic sensitivity, limited saturation power, and sub-kelvin operating requirements motivate the development of alternative nonlinear platforms. Here we demonstrate a two-mode kinetic-inductance parametric amplifier based on a pair of capacitively coupled Kerr-nonlinear resonators fabricated from NbTiN and NbN thin films. The distributed Kerr nonlinearity of these materials enables nondegenerate four-wave-mixing amplification with gains approaching \(40~\mathrm{dB}\), gain--bandwidth products up to \(6.9~\mathrm{MHz}\), and 1-dB compression powers two to three orders of magnitude higher than those of typical Josephson amplifiers. A coupled-mode model, evaluated with independently measured circuit parameters, reproduces the pump-induced modification of the hybridized modes and the measured signal and idler responses. Our results establish coupled kinetic-inductance resonators as a compact, high-dynamic-range amplifier platform, built in a material system that additionally offers the magnetic-field and elevated-temperature resilience characteristic of NbN and NbTiN, of interest for readout of superconducting qubits, spin ensembles, quantum dots, and other microwave quantum technologies.
\end{abstract}

\maketitle

Many of the most sensitive quantum-measurement platforms rely on amplifying extremely weak electromagnetic fields by adding the minimum noise allowed by quantum mechanics. Parametric amplification provides this capability by driving a nonlinear medium with a strong pump tone, redistributing pump energy coherently into a signal and its corresponding idler. With appropriate engineering of the nonlinear medium and pump conditions, this process enables high gain and low added noise approaching the fundamental quantum limit of half a photon for phase-preserving operation \cite{Caves1982,Mollow1967}. In superconducting circuits, parametric amplifiers supply the ultra-low-noise gain required to extract weak microwave signals from quantum devices \cite{Clerk2010,Aumentado2020,Blais2021,Krantz2019,Bergeal2010,Hatridge2011,Castellanos2007,bienfait2016}. They also generate nonclassical radiation, including squeezed states of light \cite{zhong2013,eichler2011,castellanos2008}, and their noise performance and phase control are critical for fast, high-fidelity, single-shot readout in superconducting qubits \cite{walter2017}, quantum dot qubits \cite{stehlik2015}, and spin ensembles \cite{vine2023,eichler2017}. Beyond qubit readout, parametric amplifiers play an important role in weak-measurement protocols \cite{Murch2013,Weber2014} and sensing \cite{PhysRevLett.118.091801,PhysRevLett.125.221302,PRXQuantum.2.040350,Backes2021, PhysRevLett.114.080503,Barzanjeh,Assouly2023,2310.07198}.

For more than a decade, Josephson junctions have served as the primary nonlinear elements for realizing quantum-limited amplifiers \cite{Yurke1985,Renger2021}, in cavity-based (JPA) \cite{yamamoto2016} and traveling-wave (JTWPA) \cite{macklin2015,Esposito} form, supporting three-wave mixing (3WM) or four-wave mixing (4WM),processes also realized with kinetic inductance \cite{malnou2021,frasca2024,mohamed2024,PRXQuantum.4.010322,khalifa2024kinetic}. Despite outstanding performance, these devices face susceptibility to magnetic parasitic noise \cite{schneider2019}, modest power-handling capability, and operational constraints that typically require sub-kelvin temperatures \cite{planat2019,Castellanos2007}. These restrictions motivate the exploration of alternative nonlinear platforms that offer increased robustness and greater fabrication simplicity.

While Josephson junction-based devices characteristically provide local, pronounced nonlinearity, superconducting thin films with strong kinetic inductance can provide extended interaction length through intrinsic distributed nonlinearity in the material \cite{mohamed2024, PhysRevApplied.22.044055, chaudhuri2017broadband,malnou2021,anferov2020,zmuidzinas2012superconducting,PhysRevApplied.16.044017}. Materials such as NbTiN \cite{mohamed2024,parker2022degenerate,khalifa2024kinetic}, NbN \cite{frasca2024,wu2024junction}, TiN \cite{joshi2022strong}, and granular aluminum \cite{zapata2024granular} exhibit a spatially uniform Kerr-type nonlinearity that enables distributed 4WM without Josephson junctions. Although the per-photon Kerr coefficient of these films is a few orders of magnitude smaller than that of a Josephson junction, the nonlinearity is distributed over the full mode volume, so that useful gain is reached at correspondingly higher pump powers while retaining a much larger dynamic range. Kinetic-inductance parametric amplifiers (KIPAs) built from these materials demonstrate high resilience to magnetic noise \cite{khalifa2023nonlinearity,parker2022degenerate}, substantially higher saturation powers \cite{mohamed2024,khalifa2024kinetic}, and operation at higher temperatures \cite{mohamed2024,10.1063/1.4931943}, making them attractive for multiplexed readout of large-scale quantum processors without extensive magnetic shielding \cite{PRXQuantum.4.010322}.

In this work, we implement and characterize a Kerr-type nonlinear microwave amplification platform based on two coupled kinetic-inductance resonators. Coupled-resonator (dimer) parametric amplifiers have been studied previously with Josephson \cite{eichler2014quantum} and kinetic-inductance \cite{zapata2024granular} nonlinearities; the contribution here is not a new conceptual framework but the realization of this architecture in NbN and NbTiN thin films and a side-by-side comparison of the two material systems using the same nominal layout.
The choice of 4WM over 3WM is a design trade-off rather than a strict advantage. A 3WM device places the pump roughly an octave from the signal, so it can be filtered before the downstream HEMT, at the cost of separate DC-bias and RF-pump lines and the noise and fabrication overhead they carry. The 4WM scheme used here needs only a single port, but the pump lies inside the signal band and is hard to reject at the output; as discussed below, this limited our noise characterization.

\begin{figure}[h!]
    \center
    \includegraphics[width=0.88\columnwidth]{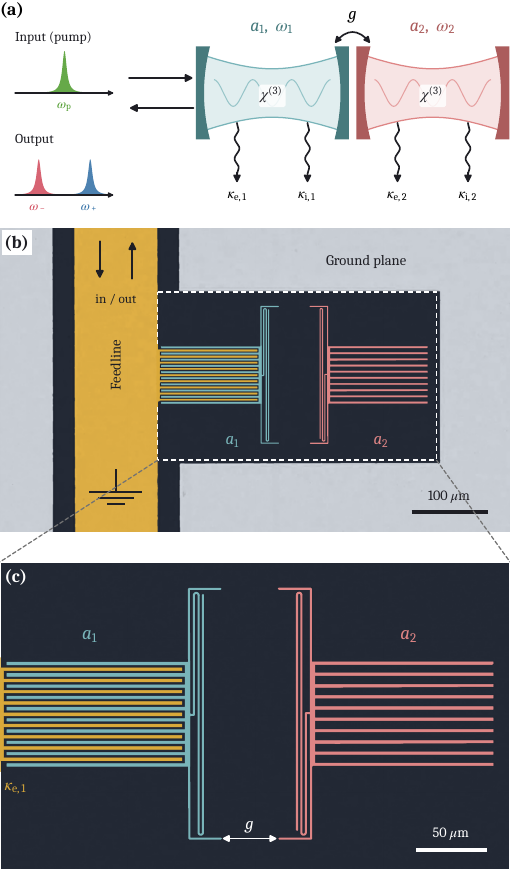}
\caption{
(a) Schematic of the coupled-resonator system used for two-mode amplification.
The device consists of two Kerr-nonlinear (\(\chi^{(3)}\)) resonators with bare
frequencies \(\omega_{1,2}\), coherently coupled at rate \(g\).
Each resonator couples to a common transmission line with external coupling rate
\(\kappa_{e,j}\) and experiences intrinsic dissipation at rate \(\kappa_{i,j}\) \((\kappa_1=\kappa_{e,1}+\kappa_{i,1})\).
The Kerr nonlinearity provides the effective \(\chi^{(3)}\) interaction required
for four-wave mixing, enabling signal and idler generation under a pump drive. (b) False-color optical image of the coupled kinetic-inductance amplifier.
The coplanar transmission line is shown in green and the ground plane in yellow.
The two nonlinear LC resonators are formed by NbTiN or NbN nanowire inductors
and U-shaped shunt capacitors and are capacitively coupled at rate \(g\). The device is driven through an interdigitated coupling capacitor connected to
the coplanar transmission line. The kinetic-inductance nonlinearity of the thin
films provides the \(\chi^{(3)}\) response responsible for 4WM
amplification.
An SEM image of the device is provided in Supplementary Sec.~I.
(c) Magnified view of the coupling region between the two resonators.
The external coupling rate \(\kappa_{e,1}\) sets the coupling of the input
field to the resonator system, while the coherent coupling rate \(g\) is controlled by the physical separation between the two resonators.\justifying}
    \label{fig:Fig1}
\end{figure}

As illustrated in Fig.~\ref{fig:Fig1}, the pump and amplified fields share a single-port reflection geometry. The feedline is terminated to ground beyond the coupling capacitor. This short-circuit termination fixes a current antinode at the coupling point, suppresses radiation into the sample box, and removes the standing-wave background of an open or through termination; its cost is a parasitic series impedance at the ground interface, minimized by dense wirebonding and accounted for in the efficiency calibration of Supplementary Sec.~II. The Kerr nonlinearity also scales quadratically with frequency (Supplementary Sec.~IV), favoring 4WM at high frequencies.
Two hybridized modes, rather than one, provide two separately engineerable branches. The resonators' bare frequencies and external couplings can be set independently, allowing the gain profile and bandwidth to be shaped without modifying the nonlinear element, while the same structure also provides a natural building block for two-mode squeezing \cite{Sivak_Shankar_Liu_Aumentado_Devoret_2020,winkel2020nondegenerate,eichler2014quantum}. The limitations should be stated equally clearly. Both modes must be engineered simultaneously, which tightens fabrication tolerances, and the two gain peaks correspond to the signal and idler of a single 4WM process rather than independent readout channels. 

Here, we summarize the coupled-mode description of two-mode amplification based on the 4WM process.
The system, shown schematically in Fig.~\ref{fig:Fig1}(a), consists of two coupled
Kerr-nonlinear (\(\chi^{(3)}\)) resonators with coupling rate \(g\).
Each resonator has a resonance frequency \(\omega_j\) and extrinsic and intrinsic
damping rates \(\kappa_{e,j}\) and \(\kappa_{i,j}\) for \(j = 1,2\), and is described by the Hamiltonian ($\hbar=1$)
\begin{equation}
    H = \sum_{j=1,2} \omega_j a_j^\dagger a_j
    + g \big(a_1 a_2^\dagger + a_2 a_1^\dagger\big)
    + \sum_{j=1,2}\frac{K^0_j}{2}\!\left[(a_j^\dagger)^2 a_j^2\right],
    \label{eq:Htot}
\end{equation}
where \(a_j\) are the annihilation operators, the second term describes coherent photon exchange at rate \(g\), and the third the self-Kerr nonlinearity with coefficient $K^0_j\corr{<0}$.
For simplicity, we assume a symmetric nonlinear regime with
\(K^0 = K^0_1 = K^0_2\). Equation~(\ref{eq:Htot}) is the nonlinear bosonic dimer, which also governs coherent Josephson-like exchange and macroscopic self-trapping between coupled cavities, exciton-polariton condensates and nanomechanical resonators \cite{Barzanjeh2016PhononJosephson, Abbarchi2013SelfTrapping, Raftery2014Dissipation}. The dimer is lithographically unbalanced, so the two branches have different external couplings ($\kappa_{e,1}\neq\kappa_{e,2}$). The self-Kerr coefficients are nonetheless set by the sheet kinetic inductance and the nanowire cross-section, common to both arms by design, and the residual asymmetry enters only through the participation ratios and bare frequencies. Since $K^0_j \propto \omega_j^{2}$ at fixed geometry and the fitted bare frequencies differ by only $2\%$, the two self-Kerr coefficients differ by $|K^0_1-K^0_2|/K^0 \simeq 4\%$, far smaller than the asymmetry of the external couplings ($\kappa_{e,1}/\kappa_{e,2} \approx 2.2$ for NbTiN and 7.7 for NbN). Relaxing the symmetric assumption and re-solving Eqs.~(\ref{SSEQM}) with $K^0_1\neq K^0_2$ at this level moves the gain maxima by less than a quarter of a linewidth and changes the peak gain by less than $0.5$~dB at fixed pump amplitude, so none of the conclusions below depend on it.

The dynamics of the system are described by the quantum Langevin equations
\begin{eqnarray}\label{EQM0}
\frac{d a_{1}}{dt} &=
-\Big[i\omega_{1} +\frac{\kappa_{1}}{2}\Big] a_{1}
-ig\, a_{2}
-i K^0\, a^\dagger_{1} a_{1}^{2} \nonumber\\
&+ \sqrt{\kappa_{e,1}}\,[\,a_p + a_{e,1}\,]
+ \sqrt{\kappa_{i,1}}\, a_{i,1},\\
\frac{d a_{2}}{dt} &=
-\Big[i\omega_{2} +\frac{\kappa_{2}}{2}\Big] a_{2}
-ig\, a_{1}
-i K^0\, a^\dagger_{2} a_{2}^{2}\nonumber\\
&+ \sqrt{\kappa_{e,2}}\,[\,a_p + a_{e,2}\,]
+ \sqrt{\kappa_{i,2}}\, a_{i,2}, \nonumber
\end{eqnarray}
where \(\kappa_j = \kappa_{e,j} + \kappa_{i,j}\) and $\kappa_{i,j}$ is the intrinsic loss rate of resonator $j=1,2$. These equations include the vacuum and thermal noise entering through the external port (\(a_{e,j}\)) and through internal loss (\(a_{i,j}\)), with the standard correlators $\langle a_{e(i)}(t) a^\dagger_{e(i)}(t')\rangle = (\bar n_{e(i)}+1)\delta(t-t')$ set by the bath occupancies $\bar n_{e(i)}$. The term \(a_p\) is the pump field. In the strong-pump regime the pump is treated classically, \(a_p \rightarrow \alpha_p = |\alpha_p| e^{-i\omega_p t}\), allowing Eqs.~(\ref{EQM0}) to be linearized as \(a_j = \alpha_j + \delta a_j\) about the steady-state amplitude \(\alpha_j\).

Transforming into a frame rotating at \( \omega_p \), the linearized
fluctuations are collected in
\(
\mathbf{A} = [\,\delta a_1,\;\delta a_1^\dagger,\;\delta a_2,\;\delta a_2^\dagger\,]^{T}.
\)
Because resonator 2 is more weakly coupled to the input line, \( \kappa_{e,1} > \kappa_{e,2} \), the terms proportional to
\( \sqrt{\kappa_{e,2}}\,[\,a_p + a_{e,2}\,] \) are neglected, so that resonator 2 is driven only through the coherent coupling \(g\). This approximation is \corr{most accurate} for the NbN device and \corr{least accurate} for NbTiN; because the pump amplitude is treated as a free parameter at each pump power, retaining the direct drive of resonator 2 mainly rescales the extracted pump amplitude. The equations of motion then read
\begin{equation}\label{EQM1}
\frac{d\mathbf{A}}{dt}
=
\mathbf{\Gamma}\,\mathbf{A}
+
\sqrt{\kappa_{e,1}}\,\mathbf{A}_{e}
+
\mathbf{A}_{i},
\end{equation}
where \(
\mathbf{A}_{e}
=
\big[\,\delta a_{e,1},\;\delta a_{e,1}^\dagger,\;0,\;0\,\big]^{T},
\) and
\(\mathbf{A}_{i}
=
\big[
\sqrt{\kappa_{i,1}}\,\delta a_{i,1},\;
\sqrt{\kappa_{i,1}}\,\delta a_{i,1}^\dagger,\;
\sqrt{\kappa_{i,2}}\,\delta a_{i,2},\;
\sqrt{\kappa_{i,2}}\,\delta a_{i,2}^\dagger
\big]^{T},
\) with

\begin{equation}\label{drift}
\mathbf{\Gamma}
=
\begin{bmatrix}
    -i\Delta_{1} -\frac{\kappa_{1}}{2} & -iK_1 & -ig & 0\\[4pt]
    iK_1^* & i\Delta_{1} -\frac{\kappa_{1}}{2} & 0 & ig\\[4pt]
    -ig & 0 & -i\Delta_{2} -\frac{\kappa_{2}}{2} & -iK_2\\[4pt]
    0 & ig & iK_2^* & i\Delta_{2} -\frac{\kappa_{2}}{2}
\end{bmatrix},
\end{equation}
in which
\( \Delta_j = \omega_j - \omega_p + 2K^0 n_j \) and
\( K_j = K^0\,\alpha_j^{2} \) is the effective Kerr coefficient set by the
intracavity photon number \( n_j = |\alpha_j|^2 \).
The amplitudes \( \alpha_j \) follow from the coupled steady-state equations

\begin{eqnarray}\label{SSEQM}
\Big[i(\omega_{1}-\omega_{p}) + \frac{\kappa_{1}}{2}\Big]\alpha_{1}
+ i g\, \alpha_{2}
+ i K^0|\alpha_{1}|^{2}\alpha_{1}
&=& \sqrt{\kappa_{e,1}}\,\alpha_p,  \nonumber\\[6pt]
\Big[i(\omega_{2}-\omega_{p}) + \frac{\kappa_{2}}{2}\Big]\alpha_{2}
+ i g\, \alpha_{1}
+ i K^0|\alpha_{2}|^{2}\alpha_{2}
&=& 0.
\end{eqnarray}

Solving Eqs.~(\ref{SSEQM}) numerically yields the steady-state photon numbers and hence \( K_j \). Multiple solutions generally exist; the physical branch is selected using the Routh--Hurwitz criterion~\cite{PhysRevA.35.5288}, retaining only steady states for which every eigenvalue of \(\mathbf{\Gamma}\) has a negative real part.

\begin{figure*}[t]
    \center
    \includegraphics[width=2\columnwidth]{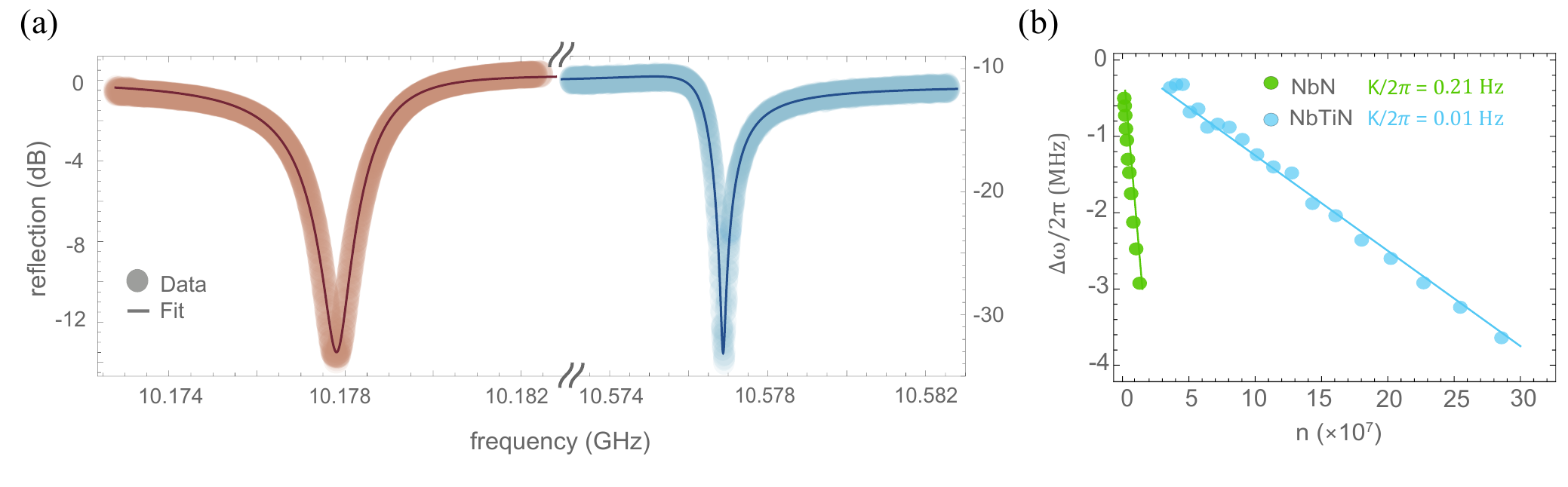}
\caption{
(a) Reflection spectrum of the NbTiN device, showing two hybridized modes at
\(\omega_{-}/2\pi = 10.178~\mathrm{GHz}\) (brown, broader linewidth) and
\(\omega_{+}/2\pi = 10.577~\mathrm{GHz}\) (blue),
corresponding to a normal-mode splitting of \(\omega_{+}-\omega_{-}=2\pi\times399~\mathrm{MHz}\); as discussed below, this equals \(2g\) only at the anticrossing.
Asymmetric coupling of the bare resonators to the single-sided feedline
results in distinct linewidths for the two normal modes. Points denote measured data, while solid lines are least-squares fits of
the complex response, including amplitude and phase, to Eq.~(S4).
The extracted parameters \(\omega_\pm\)\del{,} and \(\kappa_{j}\) are subsequently used without further adjustment to
generate the gain curves in Fig.~\ref{Fig3}(a).
(b) Self-Kerr shift as a function of intracavity photon number \(n\).
The fitted slopes using a linear function yield
\(\corr{|K|}/2\pi \approx 0.01~\mathrm{Hz}\) for NbTiN and
\(0.21~\mathrm{Hz}\) for NbN for  \(\approx 1~\mu\mathrm{m}\)-wide nanowires. The statistical uncertainty of each point, estimated from the
resonance-fit uncertainty,
is smaller than the marker size and is therefore not shown.
The dominant uncertainty in \(K\) arises from the common-mode scaling
of the photon-number axis associated with the
\(\pm 1.5~\mathrm{dB}\) input-line calibration
(Supplementary Sec.~II), yielding
\(|K|/2\pi = 0.010^{+0.004}_{-0.003}~\mathrm{Hz}\) for NbTiN and
\(0.21^{+0.09}_{-0.06}~\mathrm{Hz}\) for NbN.
\justifying}
    \label{Fig2}
\end{figure*}

Solving Eq.~(\ref{EQM1}) in the Fourier domain and inserting the result into the input--output relation
\(
a_{\mathrm{out}} = \sqrt{\kappa_{e,1}}\, \delta a_1 - a_{e,1}
\) gives
\begin{equation}\label{EqGain}
a_{\text{out}}= u(\omega) \delta a_{e,1}(\omega)+v(\omega)\delta a^\dagger_{e,1}(-\omega)+A_{\mathrm{noise}},
\end{equation}
where
\( \corr{G_{S}=|u|^2}\), \(\corr{u(\omega)=\kappa_{e,1}\,\mathcal{G}_{11}(\omega)-1} \)
and
\( \corr{G_{I}=|v|^2}\), \(\corr{v(\omega)=\kappa_{e,1}\,\mathcal{G}_{12}(\omega)}\),
with
\( \mathbf{\mathcal{G}}(\omega)=\big(-i\omega\,\mathbf{I}-\mathbf{\Gamma}\big)^{-1} \).
The operator \( A_{\mathrm{noise}} \) collects the fluctuations \( a_{i,1(2)} \) admitted by internal loss.

Equation~(\ref{EqGain}) has the standard form of a parametric amplifier \cite{Caves1982, Clerk2010}, with signal gain \(G_{S}\) and idler gain \(G_{I}\) whose spectral locations are set by the pump frequency. In the hybridized-mode picture, the coherent coupling mixes the resonators into normal modes with eigenfrequencies
\(
2{\omega}_{\pm}
=
\omega_{1}+\omega_{2}\pm\sqrt{4g^{2}+(\omega_{1}-\omega_{2})^{2}},
\)
both of which couple to the common transmission line. At the anticrossing \( \omega_{1} = \omega_{2} \) the splitting is \( \omega_{+} - \omega_{-} = 2g \). Driving midway between the hybridized modes reduces Eq.~(\ref{EqGain}) to
\(
a_{\mathrm{out}}
\propto
u\,\delta a_{e,1}(\Omega)
+
v\,\delta a_{e,1}^{\dagger}(-\Omega),
\)
with \(\Omega=(\omega_{+}-\omega_{-})/2\), which equals \(g\) only at the anticrossing,
so that amplification occurs at the normal-mode frequencies of the coupled system (all frequencies in the pump-rotating frame).

The framework above is implemented using
kinetic-inductance thin films, whose key physical mechanism is the current-dependent kinetic
inductance
\(
L_k(I) = L_0[1 + (I/I^*)^2],
\)
where \(L_0\) is the zero-bias kinetic inductance and \(I^*\) is a material-dependent scale
set by the critical current \(I_c\) \cite{Pippard1,Pippard2,Annunziata_2010,zmuidzinas2012superconducting}.
This provides the effective \(\chi^{(3)}\) nonlinearity that mediates 4WM
\cite{parker2022degenerate}.
The device incorporates two Kerr-nonlinear LC resonators
with bare frequencies \(\omega_{1,2}\), coupled capacitively at rate \(g\),
as shown in Fig.~\ref{fig:Fig1}(b) and (c). A coplanar transmission line delivers the microwave
pump to an interdigitated capacitor, which drives both resonators simultaneously.
Each mode is a compact lumped-element circuit formed by a meandered nanowire
inductor in parallel with a U-shaped shunt capacitor.

We implemented this architecture in NbTiN and NbN films, each
10\,nm thick, on high-resistivity silicon.
Fabricating identical circuit layouts in both materials enabled a side-by-side comparison of their
nonlinear response and achievable amplification at fixed layout. We note that an identical layout does not imply an identical circuit. The two films have different sheet inductances ($103$ versus $31~\mathrm{pH}/\square$), so the same geometry results in different total inductance, kinetic-inductance participation, characteristic impedance, and consequently different operating frequencies ($\sim 7.8~\mathrm{GHz}$ for NbN and $\sim 10.4~\mathrm{GHz}$ for NbTiN). The comparison below is therefore between two devices based on the same design, rather than a controlled single-variable comparison of the two materials. Where a figure of merit depends explicitly on frequency or coupling rate, we state this dependence explicitly rather than attributing the observed difference to the film alone.
The nanowire inductors and shunt capacitors
were designed to increase the kinetic-inductance participation ratio \(L_{\mathrm{K}}/L_{T}\),
thereby enhancing \(K^0\).

Figure~\ref{Fig2}(a) shows the reflection spectrum of the NbTiN device, with
two well-resolved hybridized modes at
\(\omega_{-}/2\pi = 10.178~\text{GHz}\) and
\(\omega_{+}/2\pi = 10.577~\text{GHz}\).
These parameters were extracted by fitting a linear microwave-spectroscopy model described in Sec.~III of the Supplementary Material.
The same fit for the coupled resonators returns the bare resonator frequencies and coupling, \(\omega_1/2\pi = 10.27~\mathrm{GHz}\), \(\omega_2/2\pi = 10.49~\mathrm{GHz}\) and \(g/2\pi = 166~\mathrm{MHz}\), where the index \(j=1\) labels the resonator nearer the feedline. These three quantities are not independent of the measured spectrum: they satisfy \(4g^2 = (\omega_+-\omega_-)^2-(\omega_1-\omega_2)^2\), so the quoted \(399~\mathrm{MHz}\) splitting fixes \(g\) once the bare detuning is known. We \corr{emphasize} that the normal-mode splitting equals \(2g\) only at the anticrossing \(\omega_1=\omega_2\); here the resonators are detuned by \(220~\mathrm{MHz}\), so that \(2g/2\pi = \corr{333}~\mathrm{MHz}\) while \(\omega_+-\omega_-=2\pi\times399~\mathrm{MHz}\).
Although the two resonators are nominally identical, the single-sided coupling design adds extra capacitance to the nearer resonator, which, together with fabrication variation, detunes them. This detuning arises from the single-port geometry rather than from a deliberate design choice. It does not \corr{prevent amplification}, since 4WM \corr{requires} the pump frequency to lie \corr{near the midpoint} between the two hybridized modes. The asymmetric coupling to the transmission line gives the two normal modes different linewidths,
\begin{equation}\label{eq:kappapm}
\kappa_{\pm} =
\frac{\kappa_1 + \kappa_2}{2}
\pm
\frac{\kappa_1 - \kappa_2}{2}\,
\frac{(\omega_1 - \omega_2)/g}{
\sqrt{4 + \left[(\omega_1 - \omega_2)/g\right]^2}},
\end{equation}
which conserves the total damping, \(\kappa_{+}+\kappa_{-}=\kappa_1+\kappa_2\).

\begin{figure*}[t]
    \includegraphics[width=2\columnwidth]{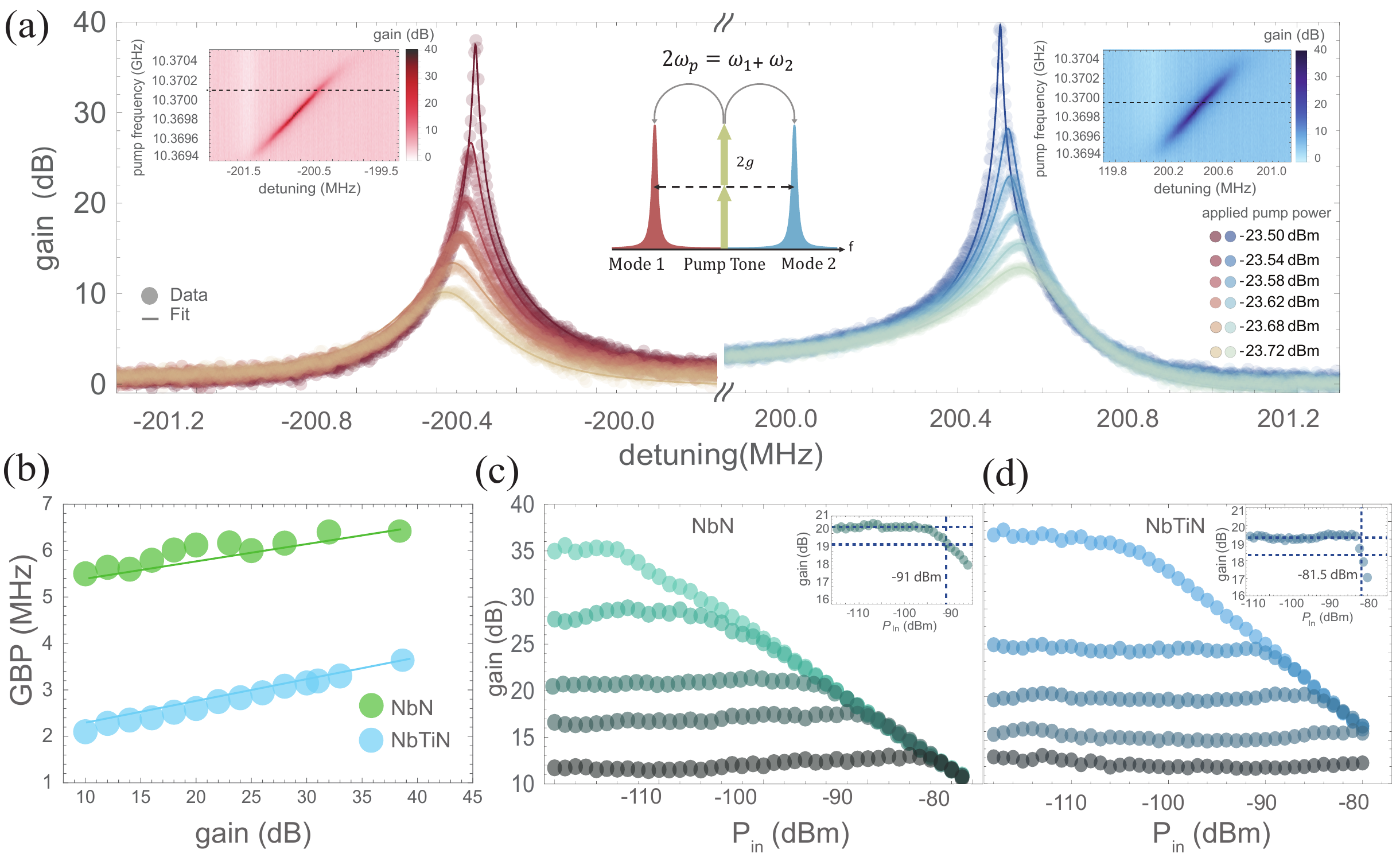}
\caption{
(a) Signal and idler gain as a function of detuning for the NbTiN device at several pump powers.
The maximum gain approaches \(40~\mathrm{dB}\) at a pump power of \(-23.5~\mathrm{dBm}\) at the device input. The gain profiles are obtained from Eq.~(\ref{EqGain}) with the device parameters fixed to the values independently extracted from the characterization measurements in Fig.~\ref{Fig2}; only the pump amplitude is adjusted at each pump power, so the lineshapes are not independently fitted. The insets show the gain landscape versus pump frequency and signal detuning, with the \corr{color} scale giving the gain in dB.
 (b) Gain--bandwidth product (GBP) as a function of gain for the NbTiN and NbN devices.
The NbTiN device reaches a maximum \(\mathrm{GBP}=3.3~\mathrm{MHz}\), while the NbN device reaches
\(\mathrm{GBP}=6.9~\mathrm{MHz}\); as discussed in Supplementary Sec.~V, the gain--bandwidth product is set by the normal-mode linewidths and not by the Kerr coefficient. The bandwidth (BW) is obtained by fitting each measured gain profile in panel (a) \corr{(NbTiN) and Fig.~S4(c) (NbN)} to a Lorentzian function and extracting its full width at half maximum (FWHM).
The gain--bandwidth product is then calculated as
\(\mathrm{GBP}=\sqrt{G}\,\mathrm{BW}\), where \(G\) is the linear power gain.
 (c,d) Power-saturation characteristics of the NbTiN and NbN devices, respectively.
At \(20~\mathrm{dB}\) gain, the input-referred 1-dB compression points are
\(P_{\mathrm{1dB}}=-81.5~\mathrm{dBm}\) for NbTiN and
\(P_{\mathrm{1dB}}=-91~\mathrm{dBm}\) for NbN, corresponding to \(-61.5\) and \(-71~\mathrm{dBm}\) referred to the device output.
\(P_{\mathrm{1dB}}\) is defined as the input power at which the small-signal gain has fallen by \(1~\mathrm{dB}\); the dashed lines are the corresponding construction lines.
The roll-off beyond compression is steeper than \(1~\mathrm{dB}\) per \(1~\mathrm{dB}\) of input power, as expected for a Kerr-limited resonant amplifier, in which the intracavity population pulls the mode away from the pump and the gain collapses over a narrow power range rather than compressing gradually. 
\justifying}
    \label{Fig3}
\end{figure*}

The extracted damping rates for NbTiN are
\corr{\(\kappa_{1}/2\pi=2.3~\mathrm{MHz}\)} and \corr{\(\kappa_{2}/2\pi=1.21~\mathrm{MHz}\)},
and for NbN
\corr{\(\kappa_{1}/2\pi =4.06~\mathrm{MHz}\)} and \corr{\(\kappa_{2}/2\pi =2.16~\mathrm{MHz}\)}. All four rates are fitting parameters extracted from the complex-$S_{11}$ fits shown in Fig.~\ref{Fig2}(a) and Fig.~S4(a), rather than from electromagnetic simulations. The fitting model, constraints, and uncertainties are described in Supplementary Sec.~III. Inserting these rates and the bare parameters above into Eq.~(\ref{eq:kappapm}) gives normal-mode linewidths $\kappa_-/2\pi = \corr{2.06}~\mathrm{MHz}$, $\kappa_+/2\pi = \corr{1.45}~\mathrm{MHz}$ and $\kappa_-/2\pi = 3.87~\mathrm{MHz}$, $\kappa_+/2\pi = 2.3~\mathrm{MHz}$ for NbTiN and NbN respectively, 
 whose sum equals $\kappa_1+\kappa_2$ and which reproduce the observation that the lower hybridized mode is the broader of the two.

For a lumped resonator coupled through \(C_c\) to a transmission line of impedance \(Z_0\),
\(\kappa_e=\omega^2 C_c^2 Z_0/C_\Sigma\) \cite{Goppl2008} where $C_\Sigma$ is the total circuit capacitance.
Since \(C_\Sigma\) and \(C_c\) are set by the common mask geometry, identical layouts predict
\(\kappa_e\propto\omega^2\), so the NbN device should be more weakly coupled by
\((7.84/10.38)^2=0.57\). Instead, we measure
\(\kappa_{e,1}^{\mathrm{NbN}}/\kappa_{e,1}^{\mathrm{NbTiN}}=1.6\),
corresponding to an effective \(C_c\) larger by a factor of \(1.7\).
We attribute this difference to fabrication rather than material properties:
the SEM in Fig.~S1(c) shows metal features about \(20\%\) wider than designed,
which narrows the coupling-capacitor gaps and increases \(C_c\).
We therefore treat the difference in \(\kappa_e\) as a chip-to-chip lithographic variation and make its impact explicit in the device comparison of Supplementary Sec.~V.
The internal losses show no additional penalty from the coupled-resonator geometry:
\(\kappa_i/2\pi=0.65\)--\(1.1~\mathrm{MHz}\) across the four modes, corresponding to
\(Q_i=0.7\)--\(1.6\times10^4\), consistent with uncoupled nanowire resonators fabricated from the same films and process \cite{mohamed2024,frasca2024}.
The architecture-specific parasitic loss arises from the shorted feedline termination discussed above; dense ground wirebonding minimizes this contribution, and its residual effect is included in the measurement-efficiency calibration of Supplementary Sec.~II.

To determine the Kerr coefficient, we measure the resonance shift versus
intracavity photon number, whose slope gives
\(K_j = \Delta\omega_j / n_j\).
As shown in Fig.~\ref{Fig2}(b), \(\corr{|K|}/2\pi \approx 0.01~\text{Hz}\) for NbTiN and
\(0.21~\text{Hz}\) for NbN with
1\,$\mu$m-wide nanowires. The expected scaling,
\(K \propto \omega^{2}(L_{\mathrm{K}}/L_T)^{2}/(N_0\Delta_0^{2}V)\)~\cite{joshi2022strong,frasca2023nbn}, contains both circuit and material factors. Thus for the common layout used here the circuit terms account for a factor of two of the observed twenty-fold difference, the remainder following from the material prefactor (Supplementary Sec.~V). Here $L_K=L_k(0)$ and $L_T$ is total linear inductance.

To observe amplification, the circuit is driven with a pump tone placed near the midpoint of the
two hybridized modes, \(\omega_{p} \simeq (\omega_{+} + \omega_{-})/2\), activating a 4WM process that converts
pump photons into signal and idler fields at the two normal modes; the device then operates as a nondegenerate, phase-insensitive
amplifier. For NbTiN, Fig.~\ref{Fig3}(a) shows the measured signal and idler gain versus detuning
at different pump powers, reaching nearly \(G=40~\mathrm{dB}\) at
\(\corr{-23.5}~\mathrm{dBm}\) at the device input. The solid curves are Eq.~(\ref{EqGain}) evaluated with the independently extracted parameters of Fig.~\ref{Fig2}(a), with only the pump amplitude adjusted at each pump power. The insets display the amplification landscape versus pump frequency and detuning for signal and idler. The NbN device reaches comparable performance at \(-36~\mathrm{dBm}\) (Supplementary Sec.~VI); Table~S1 of the Supplementary Material compares the two devices. Figure~\ref{Fig3}(b) presents the gain--bandwidth product
\(\mathrm{GBP} = \sqrt{G}\,\Gamma\), with \(\Gamma\) the FWHM of the gain profile:
\(3.3~\mathrm{MHz}\) for NbTiN and
\(6.9~\mathrm{MHz}\) for NbN near 40~dB gain. Theoretically \(\sqrt{G}\,\Gamma \sim \kappa_{+}\kappa_{-}/(\kappa_{+}+\kappa_{-})\)
\cite{Clerk2010}, up to a numerical factor of order unity that depends on the pump detuning and on the hybridization angle, which identifies the regime in which the GBP can be optimized at fixed gain.

Gain saturation occurs when the Kerr-induced intracavity frequency shift becomes comparable to the mode linewidth.
Figures~\ref{Fig3}(c)--(d) show the measured gain versus input probe power.
The probe power is referred to the device input plane using the calibrated input attenuation and cable loss (Supplementary Sec.~II).
For each probe power, spectra were recorded with the pump ON and OFF, and \(P_{\mathrm{1dB}}\) is defined as the input power at which the small-signal gain decreases by \(1~\mathrm{dB}\).

Using the saturation condition \(Kn\sim\kappa\) together with
\(n=4\kappa_e P_{\mathrm{in}}/(\hbar\omega\kappa^2)\) gives 
\(
P_{\mathrm{1dB}}
\propto
\frac{\hbar\omega\,\kappa^3}{\kappa_e K},
\) \cite{eichler2014controlling}
showing that saturation depends jointly on the Kerr coefficient and linewidth.
Using the independently measured \(\omega\), \(\kappa\), \(\kappa_e\), and \(K\), this scaling numerically predicts the NbN device to saturate \(11.6~\mathrm{dB}\) below the NbTiN device.
The \(21\times\) larger Kerr coefficient of NbN is partly offset by its larger linewidth, while its lower resonance frequency acts in the same direction as the Kerr coefficient.
Experimentally, we obtain
\(P_{\mathrm{1dB}}=\corr{-81.5}~\mathrm{dBm}\) for NbTiN and
\(-91~\mathrm{dBm}\) for NbN at \(20~\mathrm{dB}\) gain, both referred to the device input.
The measured \(9.5~\mathrm{dB}\) difference agrees in sign and scale with the prediction; the remaining discrepancy is consistent with the combined input-line calibration uncertainty and the gain-dependent prefactor \(\eta(G)\) discussed in Supplementary Sec.~V.
The larger-Kerr NbN device therefore saturates at lower input power, as expected. Output-referred, the compression powers are \(-61.5\) and \(-71~\mathrm{dBm}\) for NbTiN and NbN, respectively.
The NbTiN value is comparable to other resonant KIPAs with the same bandwidth
\cite{khalifa2023nonlinearity}.

The compression point is set by Kerr-induced detuning rather than the film critical current.
Writing
\[
\omega(I)\simeq
\omega_0
\left[
1-\frac{p}{2}\left(\frac{I}{I^{*}}\right)^2
\right],
\]
with \(p=L_K/L_T\), the condition \(Kn\sim\kappa\) gives
\[
\left(\frac{I}{I^{*}}\right)^2
=
\frac{2\kappa}{p\omega_0}.
\]
Using \(p=0.32\) for NbTiN and \(p=0.61\) for NbN
(Supplementary Sec.~V) gives \(I/I^{*}\simeq0.037\) for both devices.
Since \(I^{*}\) and the depairing current \(I_c\) differ only by a factor of order unity,
the intracavity current at compression stays below \(10\%\) of \(I_c\).
Higher saturation power therefore follows primarily from balancing Kerr nonlinearity and linewidth rather than increasing \(I_c\).

Evaluating this architecture requires weighing several trade-offs. NbN and NbTiN provide high kinetic inductance, hence a compact footprint and high characteristic impedance, and their large critical-current density gives a wide dynamic range. Conversely, they require a strong pump tone and, being disordered superconductors, contribute additional dielectric loss. We finally note that the pump is applied at the mean frequency of the hybridized modes, so a significant fraction of it reflects toward the input port and saturates the subsequent amplification chain. We implemented a pump-cancellation scheme \cite{eichler2013experimental}, suppressing the pump by destructive interference at a directional coupler, but the achievable isolation was insufficient to keep the HEMT out of saturation, and the cancellation had to be re-tuned at every pump power. This prevented a reliable determination of the added noise, and we therefore make no quantum-limited claim for these devices; a dedicated noise characterization, most naturally \corr{in a 3WM version} of the same dimer in which the pump can be filtered, is left to future work.

We have demonstrated a coupled-mode kinetic-inductance parametric amplifier providing
two-mode, nondegenerate gain through 4WM in NbTiN and NbN
nanowire resonators, with gain approaching
\(40~\mathrm{dB}\), megahertz-scale gain--bandwidth products, and saturation powers
orders of magnitude above Josephson-junction amplifiers.
The NbN device exhibits a larger Kerr coefficient and a larger
gain--bandwidth product, at the cost of a correspondingly lower saturation power. Its higher kinetic-inductance participation and lower operating frequency account for only a factor of two of the twenty-fold difference in Kerr coefficient, while the gain--bandwidth product is set by the normal-mode linewidths and hence by the coupling design rather than by the film. Because the same layout in the two films yields different participation ratios ($0.61$ versus $0.32$) and different operating frequencies, these differences are not attributable to the choice of film alone; Supplementary Sec.~V separates the circuit-parameter and material contributions explicitly. Our coupled-mode theory captures the pump-induced modification of the hybridized modes and the emergence of signal and
idler tones at the \corr{lower and upper hybridized modes}, providing clear guidance
for optimizing future device geometries.

Looking ahead, raising the kinetic-inductance participation---for example with
narrower nanowires or materials with stronger nonlinearity such as TiN or granular
\corr{aluminum}---and reducing internal losses should further enhance the Kerr nonlinearity and lower the required drive power.
Traveling-wave kinetic-inductance amplifiers (KI-TWPAs) achieve multi-gigahertz instantaneous bandwidths using long nonlinear transmission lines \cite{ho2012wideband,malnou2021,faramarzi2025near}, at the cost of \corr{centimeter}-scale footprints, dispersion engineering, and high pump powers. Resonator-based KIPAs offer compact footprints and lower pump requirements but are constrained by the gain--bandwidth product; impedance-engineered matching networks \cite{hung2025broadband} and hybridized multi-resonator coupling---such as the dimer demonstrated here---provide routes around this limit.
The present device cannot compete with KI-TWPAs in instantaneous bandwidth. Our $6.9~\mathrm{MHz}$ GBP is orders of magnitude smaller than the multi-GHz bandwidths reported in Refs.~\cite{ho2012wideband,malnou2021}, and also below that of impedance-engineered KIPAs \cite{hung2025broadband}. The dimer instead offers a complementary architecture with a \corr{sub-millimeter} footprint, no dispersion-engineering features or transformer sections, and single-step lithography. Cascading additional coupled modes or terminating the dimer with a matching network \cite{hung2025broadband}, following the approach established for impedance-matched Josephson amplifiers \cite{Mutus,macklin2015}, provides a natural route toward broader bandwidth while retaining a comparatively simple architecture.
Resilience to
magnetic noise, high saturation power, and elevated-temperature operation ---platform properties established in earlier NbN/NbTiN work \cite{parker2022degenerate,khalifa2023nonlinearity,vine2023,mohamed2024} rather than measured here--- make such amplifiers attractive
for multiplexed readout in superconducting-qubit, spin, and quantum-dot platforms.

We acknowledge funding by the Natural Sciences and Engineering Research Council of Canada (NSERC) through its Discovery Grant and Quantum Alliance Grant, and NSERC through Advance Grant project, and Alliance Quantum Consortium. This project is funded in part by the Government of Canada. Ce projet est financé en partie par le gouvernement du Canada.

\textbf{Author contributions:}
D.D. and A.M. fabricated the devices. D.D. designed and performed the measurements. D.D. and S.B. analyzed the data and developed the theory. S.B. conceived and supervised the project. D.D. and S.B. wrote the manuscript\corr{.}

\bibliography{References}

\end{document}


\title{Supplementary Materials for "Hybridized-Mode Parametric Amplifier in Kinetic-Inductance Circuits"}

\maketitle

\section{Fabrication Recipe }
\begin{figure}[b]
\includegraphics[width=11cm]{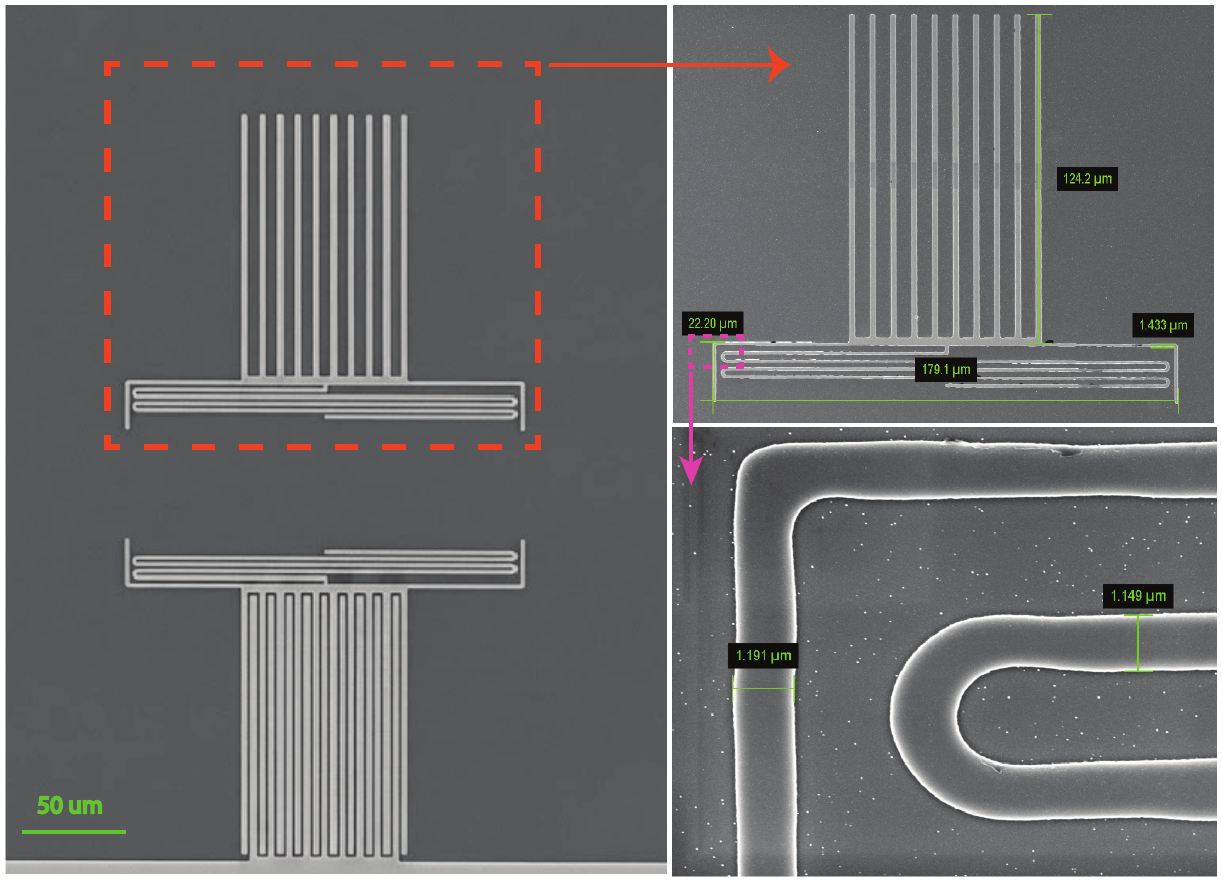}
\caption{
SEM images of the coupled-resonator device fabricated from NbTiN.
(a) Overview of the complete coupled-resonator structure.
(b) Close-up of a single resonator, where the meandering nanowire provides the inductive element and the U-shaped structure forms the shunt capacitance.
(c) Magnified view of the nanowire width. The fabricated nanowire is approximately \(20\%\) wider than the nominal layout value, leading to a shift in the resonance frequency and a reduction of the Kerr nonlinearity. ALT TEXT: Scanning electron microscopy images showing the overview and magnified details of a coupled-resonator device fabricated from a niobium-titanium-nitride thin film. The close-up panels highlight the meandered nanowire inductor and U-shaped shunt capacitor components. 
 \justifying }
\label{figSEM}
\end{figure}

The devices are fabricated using a straightforward single-step optical-lithography process.
A ten-\corr{nanometer} superconducting thin film of niobium–titanium–nitride or niobium–nitride
is deposited on a five-hundred–\corr{micrometer}, high-resistivity silicon wafer by atomic-layer
deposition. A protective layer of polymethyl methacrylate (PMMA A8) is applied to shield
the film during dicing into ten-by-ten–\corr{millimeter} chips. Each chip is first cleaned in
acetone and isopropyl alcohol in a sonication bath, then prebaked and coated with
AZ1529 photoresist, producing a film thickness of approximately three micrometres before
a soft bake at one hundred degrees Celsius. The patterns are written using a maskless
optical lithography system, after which the exposed regions are developed in AZ~400K
solution diluted at a ratio of one to four and subsequently rinsed in deionized water.
The structures are then transferred into the superconducting film by a six-minute
etch in an inductively coupled plasma reactor using \corr{sulfur} hexafluoride, argon, and
trifluoromethane gases. The 3~\(\mu\)m of photoresist is sufficiently thick to survive the etch, and the 6-minute total duration ensures complete film clearance in the narrow gaps, at the price of an overetch into the silicon substrate, which also helps mitigate interface dielectric loss \cite{bruno2015reducing,calusine2018analysis}. 
Finally, the chips are exposed to ultraviolet light to remove
any residual resist and cleaned again in acetone to complete the process. After fabrication, the chips are mounted in an oxygen-free, high-conductivity copper
sample enclosure using silver paste to ensure good thermal anchoring to the ten–millikelvin
stage. Microwave connections are made by bonding the on-chip launch pads, designed with
fifty-ohm impedance, to the printed circuit board using aluminum wire bonds and surface-mount
connectors. Additional wire bonds between the on-chip ground planes and the printed circuit
board ground are added to eliminate charge build-up and to ensure stable electrical
performance during cryogenic operation. The same dense ground bonding also minimizes the parasitic series impedance of the short-circuit feedline termination discussed in the main text. \corr{A zoomed-in} image of the sample is shown in Fig. \ref{figSEM}.

\begin{figure}[t]
\includegraphics[width=14.5cm
]{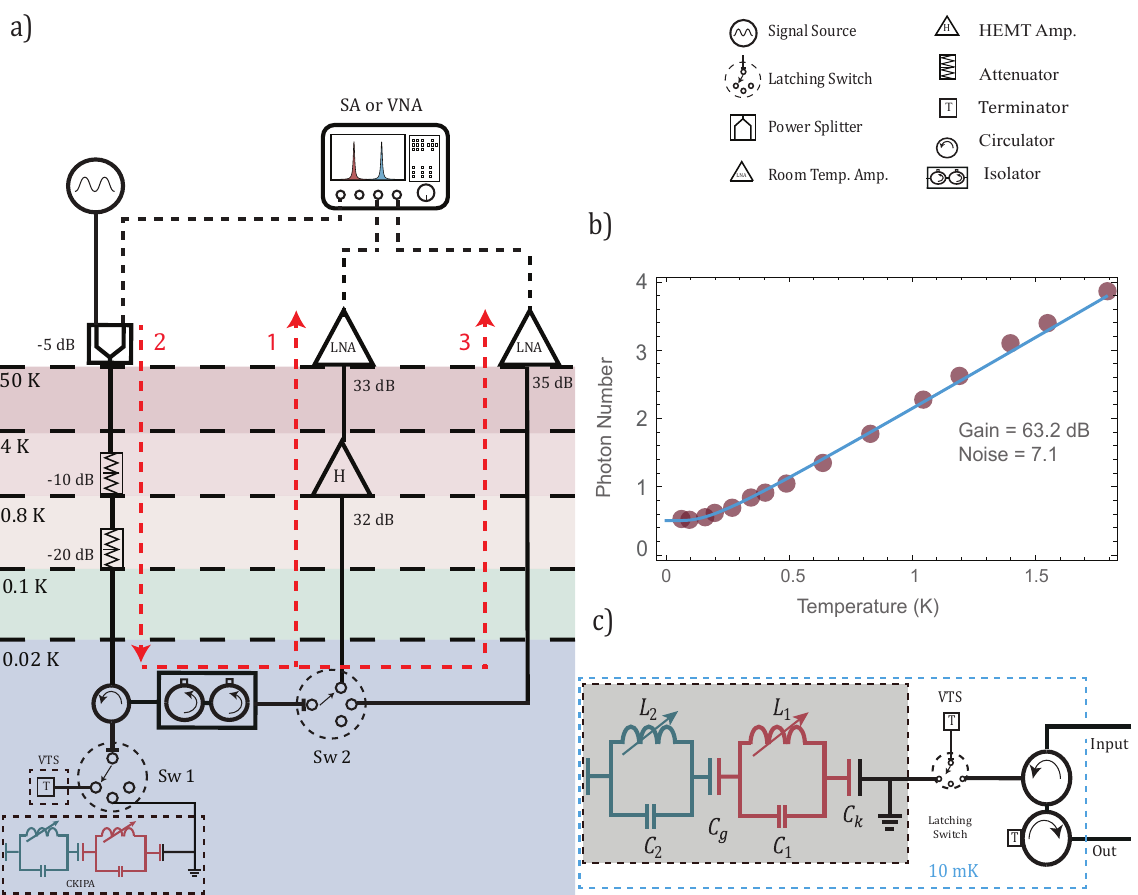}
\caption{
(a) Microwave measurement setup inside the dilution refrigerator. The system uses a single input line and switchable access to two output ports. A coherent tone or VNA signal is sent through input line~2, reflected from the device connected to switch~2, and routed to either output~1 or output~3 depending on the switch position. A variable-temperature source (VTS) is used to calibrate output~1 and determine the gain and added noise of the measurement chain. (b) Calibration of output~1 for gain and added-noise measurements. 
The measured output noise is expressed as an input-referred equivalent photon number, \(N=N_{\mathrm{out}}/G=P_{\mathrm{out}}/(\hbar\omega\,\mathrm{BW}\,G)\), and plotted versus the physical temperature of the \(50~\Omega\) VTS load according to Eq.~(S1). The solid line shows a fit to Eq.~(S1), from which the total output chain gain and system-added noise are extracted. The horizontal axis is the physical VTS temperature in kelvin, whereas the vertical axis is the dimensionless noise photon number. Statistical uncertainties are smaller than the marker size: each point is obtained from \(500\) averaged spectra in a \(100~\mathrm{Hz}\) bandwidth, giving a standard error below \(1\%\). The VTS thermometry uncertainty is \(\pm5~\mathrm{mK}\) and is negligible above \(200~\mathrm{mK}\). A weighted least-squares fit over \(50\)--\(1700~\mathrm{mK}\) yields a total output-chain gain \(G=63.2\pm0.3~\mathrm{dB}\) and a system added noise \(n_{\mathrm{add}}=7.1\pm0.4\), the quoted intervals being the \(1\sigma\) confidence intervals obtained from the fit covariance matrix. The residuals show no systematic structure, indicating good thermalization of the load over the full temperature range. The complete uncertainty budget is given in Sec.~VII. (c) Measured coherent-tone power at output~1 (red) and output~3 (blue). Output~1 contains an additional amplifier and therefore shows a higher detected power. 
The gaps of capacitors \(C_1\) and \(C_2\) are below the resolution of the optical image and are resolved in the SEM image of Fig.~\ref{figSEM}(b). ALT TEXT: A schematic diagram and scatter plots detailing the dilution refrigerator microwave measurement setup and calibration results. The graphs display the extracted total output-chain gain, and system-added noise for our measurement output lines. \justifying}
\label{figsetup}
\end{figure}
\section{Setup design and intracavity photon number}
\label{sec:calib}
Measuring the correct number of photons in the cavity requires the accurate calibration of multistage cryo-amplifiers (HEMT) and room-temperature low-noise amplifiers (LNA). The Friis formula for the noise factor (\(\text{n}_{add}\)) facilitates calculating accurate added noise in multistage amplifiers, cryogenic components, and a radial latching microwave switch (\corr{Radiall} R573 423 600). Desired spectral noise density at \(\omega/2\pi=10.187\) GHz was introduced to the output line 1 (HEMT + LNA) by 50 \(\Omega\) cryo-terminators mildly anchored with three copper braids to the mixing chamber at approximately 10 mK. A Variable Temperature Source (VTS) composed of a low-current DC heater, cryo-terminators, and  temperature sensor was thermalized carefully to generate desired thermal states. \corr{The VTS temperature was swept from 50 mK to 1700 mK while the noise power was measured} with a spectrum analyzer (\corr{Rohde \& Schwarz} FSW) in \corr{a} bandwidth (BW) of 100 Hz\corr{,} averaging 500 \corr{traces}. The following equation explains the relation between noise power and temperature, considering frequency \(\omega\), total gain G, introduced noise \(n_{add}\).
\begin{equation}
\begin{split}
N_{out} = \frac{P_{out}}{\text{BW}\hbar\omega} = G\left[\frac{1}{2}\text{coth}\left(\frac{\hbar\omega}{2k_{B}T}\right)+n_{add}\right]
\end{split}
\end{equation}

In Fig.~\ref{figsetup}(b), the experimental data and the fitted plot are shown as dots and solid lines, respectively. Firstly, we accurately extract the effective gain \(G=63.2\pm0.3~\mathrm{dB}\) and the number of added noise photons in the output line \(n_{\mathrm{add}}=7.1\pm0.4\) near the first mode.
Secondly, the effective attenuation in the input line \corr{2} needs to be measured. Validating the precise amount of attenuation allows us to determine the input line attenuation (\(P_{A}\)) and calculate the exact delivered input power at the device port. Therefore, the aforementioned gain and noise values correspond to output 1 in Fig.~\ref{figsetup}(a). The extracted gain value enables us to determine the precise effective attenuation ($A=-36.5\pm0.3$ dB) of input line 2, from the first connector to switch 1, which is required for calculating the photon number within the cavity. This value was obtained by sending a coherent tone and measuring the output power using the spectrum analyzer (SA), as shown in Fig.~\ref{figsetup}(c) with red dots. Summing the gain, input power, and measured source power values yields the total attenuation of input line 2.

Lastly, switch 2 was redirected to output 3, and measurements were performed in the same manner as in step 2. The signal was reflected from switch 1 and measured through output 3. By knowing the attenuation and utilizing both the coherent tone input and the measured value at the SA port, we determined the gain of output 3. However, due to the removal of the HEMT amplifier on output 3—necessitated by its failure under a strong pump tone—measuring thermal noise in the range of 50 mK to 1700 mK through output 3 was no longer possible, resulting in inaccurate values.\corr{ }Furthermore, input-output theory elaborates on the number of photons in a cavity; writing the field equation of motion for a single cavity under a classical pump tone yields:\begin{equation}\begin{split}n_{cavity}= \frac{P_{\text{W}}}{\hbar\omega_{0}}\frac{4\kappa}{(\kappa+\gamma)^2+4\delta^2}.\end{split}\end{equation}Here, $P_{\text{W}}=10^{(\frac{P_{t}+P_A-30}{10})}$ represents the signal power at the device input. This parameter dictates the number of photons in the cavity, where $P_t$ is the driving pump power, and $P_A$ is the attenuation of the input line at the resonance frequency $\omega_0$. The system is characterized by an external coupling rate $\kappa$, an internal loss rate $\gamma$, and a detuning $\delta$ \cite{Fink2016}. All input powers quoted in the main text, including the 1-dB compression points, are referred to this plane, i.e.\ to the device input, with the line attenuation and cable loss subtracted at the relevant amplifier frequency; the residual calibration uncertainty is $\pm1.5$~dB, dominated by the temperature dependence of the attenuator chain between the calibration reference plane and the device, and by the frequency interpolation of the cable loss between the two amplifier bands. The full budget is given in Sec.~VII. The determined power range and the corresponding frequency shifts are shown in Fig.~\ref{figure33}(b). As previously noted, due to the saturation of the output line, we removed the HEMT in output 3 and performed the measurements using only the room-temperature amplifier. This configuration did not allow us to measure the gain and noise of output 3 directly; therefore, we employed an indirect method to calculate the output gain. First, we calibrated the gain and noise of output line 1 using the VTS and the fitting formula provided in Eq.~(S1). Following this, we used a coherent tone to measure the actual attenuation of input 2, subsequently applying the same method to determine the actual gain of output 3.

\subsection*{Pump leakage and attempted pump cancellation}

Because the pump lies midway between the two hybridized modes, it falls within the amplification band and cannot be removed by filtering. We therefore implemented an analog pump-cancellation interferometer \cite{eichler2013experimental}: a copy of the pump, with independently controlled amplitude and phase, was injected through a \(20~\mathrm{dB}\) directional coupler on the output line to destructively interfere with the reflected pump. At fixed pump power, this provided approximately \(20~\mathrm{dB}\) of suppression, which was insufficient to prevent the residual pump from driving the HEMT into compression.

The required suppression is substantial. For a pump power of \(-23~\mathrm{dBm}\) at the device input, the nearly fully reflected pump reaches the HEMT at a level comparable to its \(1~\mathrm{dB}\) compression point after accounting for the output-line attenuation. Approximately \(46~\mathrm{dB}\) of cancellation is therefore required to keep the HEMT safely below compression, about \(\corr{26}~\mathrm{dB}\) beyond the suppression achieved with our single-stage analog interferometer.

The optimal cancellation amplitude and phase also drifted with pump power, requiring re-optimization at each operating point. This made systematic added-noise measurements across the gain curve impractical. We therefore report no added-noise value for these devices and make no claim of quantum-limited operation. A natural route toward reliable noise characterization is a three-wave-mixing version of the same dimer, where the pump is spectrally separated from the signal and idler and can be efficiently filtered. This implementation is the subject of ongoing work.

\begin{table*}[t]
\caption{Comparison of device performance. Saturation powers are referred to the device input plane at \(G = 20~\mathrm{dB}\); the corresponding output-referred values are 20~dB higher. Literature values have been converted to the same convention where necessary. ALT TEXT: A data table comparing the key performance figures of merit across different parametric amplifier platforms. It contrasts parameters like gain-bandwidth product, saturation power, and self-Kerr coefficients for various superconducting devices. 
}
\begin{ruledtabular}
\begin{tabular}{lcccc}
\textrm{Figure of Merit} &
\textrm{NbN} &
\textrm{NbTiN} &
\textrm{gr-Al \cite{zapata2024granular}} &
\textrm{SQUID \cite{eichler2014quantum}} \\
\colrule
Gain Bandwidth Product & 6.9 MHz & 3.3 MHz & 28 MHz & 275 MHz \\
Operating frequency \(\bar\omega/2\pi\) & 7.84 GHz & 10.38 GHz & 8.19 GHz & 7.41 GHz \\
Saturation point at $G = 20\,\mathrm{dB}$ (input-referred) & -91 dBm & \corr{-81.5 dBm} & -111 dBm & -111 dBm \\
Saturation point, output-referred & -71 dBm & \corr{-61.5 dBm} & -91 dBm & -91 dBm \\
Self-Kerr $K_0/2\pi$ coefficient & 0.21 Hz & 0.01 Hz & 1.3 kHz & 80 kHz\\
Pump power  & -36 dBm & \corr{-23.5 dBm} & -67 dBm & -75 dBm\\
Intrinsic loss \(\kappa_{i}/2\pi\) & 0.9--1.1 MHz & 0.65--0.91 MHz & < 4.4 MHz & - \\
Sheet inductance & 103 pH/$\Box$ & 31 pH/$\Box$  & 120 pH/$\Box$ & -- \\
Kinetic-inductance participation \(L_K/L_T\) & 0.61 & 0.32 & 0.80 & -- \\
Normal-mode splitting \(\omega_+-\omega_-\) & 263 MHz & 399 MHz & 198 MHz & 380 MHz \\
Exchange coupling ($g/2\pi$) &  99 MHz & 166 MHz & 99 MHz & 250 MHz \\
Nanowire cross-section & 0.01 $\mu m^2$ & 0.01 $\mu m^2$ & 0.008 $\mu m^2$ & - \\
Critical current \(I_c\) & $\sim0.1$--$1$ mA (est.) & $\sim0.1$--$1$ mA (est.) & -- & -- \\
\(I/I_c\) at \(P_{\mathrm{1dB}}\) & $<0.1$ & $<0.1$ & -- & -- \\
\end{tabular}
\end{ruledtabular}
\label{tab:comparison}
\end{table*}

\begin{figure}[b]
\includegraphics[width=16cm]{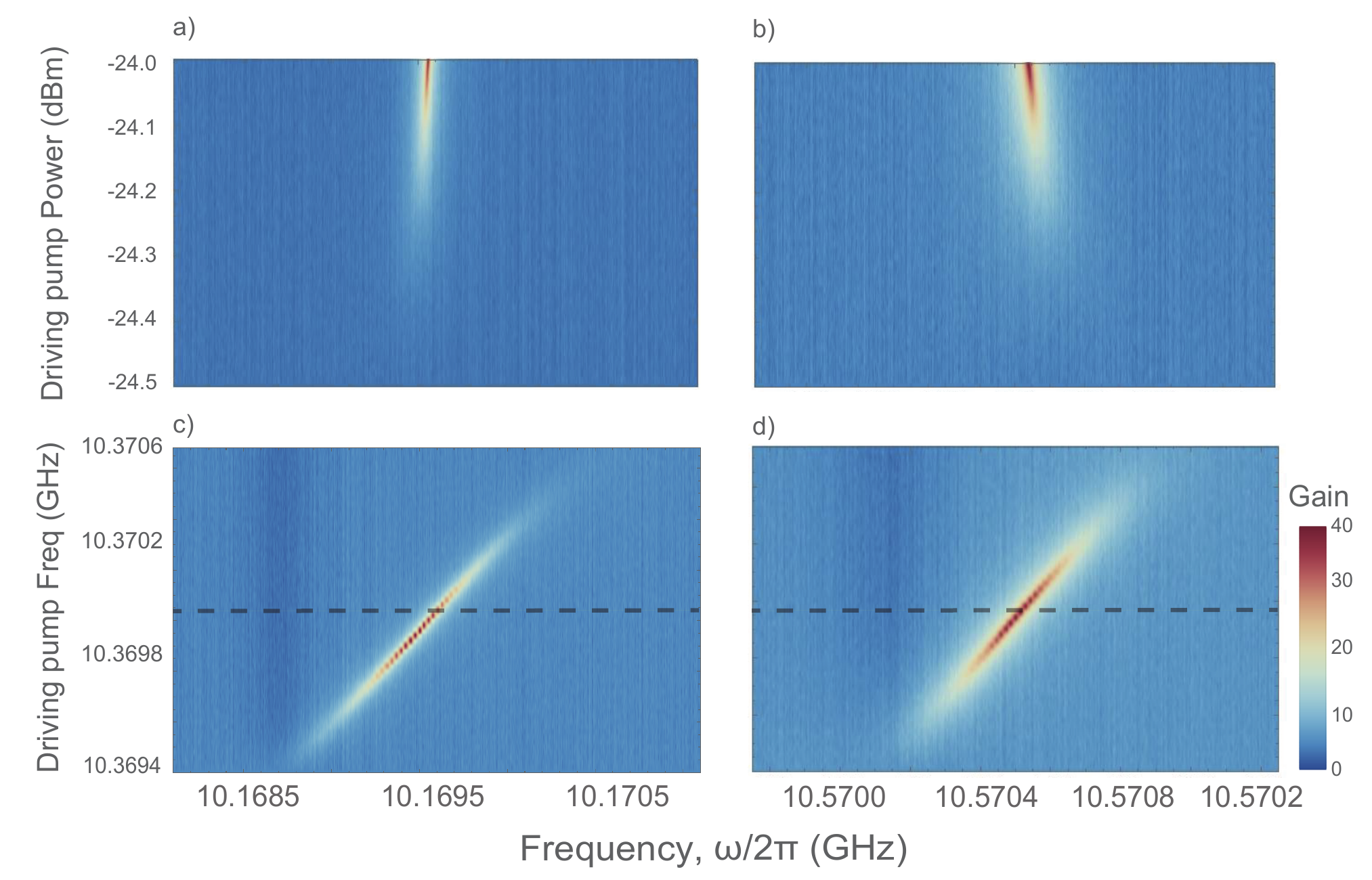}
\caption{(a,b) Pump-power sweeps at the device input plane for the two hybridized modes. The \corr{color} scale is gain in dB; the red region marks the onset of high gain once the pump exceeds threshold. (c,d) Pump-frequency sweeps showing the condition for two-mode amplification: energy conservation and photon-pair generation require the pump to sit near the midpoint of the two modes, indicated by the black dotted line. \corr{These} four panels define the operating point of the parametric amplifier. ALT TEXT: Spectral heat maps illustrating pump-power and pump-frequency sweeps at the device input plane for two hybridized modes. These parameter sweeps identify the optimal operating point for achieving high two-mode parametric amplification.
\justifying}
\label{figure44}
\end{figure}

\section{Mode Characterization and Resonator Nonlinear Response}
Operation of parametric amplifiers with resonators requires a precise characterization of their modes. Assuming a weak drive, for which the Kerr nonlinearity is negligible compared with the resonator linewidth, we analyze the frequency response of the device when the transmission line is shorted to ground. The topology corresponds to a reflection-type single-port configuration in which the signal \corr{traveling} down the feedline is reflected back:
\begin{equation}
    \underbrace{A e^{-i(2\pi\nu\tau+\phi)}}_{\text{Line / Environment Factor}} \times \underbrace{\left[ 1 - \frac{Q_i - i Q_i Q_e \frac{\delta\nu}{\nu_0}}{(Q_i + Q_e) + 2i Q_i Q_e \frac{\nu - \nu_0}{\nu_0}} \right]}_{\text{\corr{Lorentzian} lineshape}}
\end{equation}

Here the resonance is characterized by the frequency \(\nu_0\) and the quality factor \(Q = \nu_0/\kappa = Q_e Q_i/(Q_e + Q_i)\), with \(Q_e = \nu_0/\kappa_e\) the extrinsic quality factor (\(\kappa_e\) the external coupling) and \(Q_i = \nu_0/\kappa_i\) the intrinsic quality factor (\(\kappa_i\) the internal loss). The global amplitude \(A\), added phase \(\phi\) and delay \(\tau\) (accounting for cables, switches and connectors) are used to fit the uncalibrated reflection signal, and the impedance mismatch of the input line is captured by \(\delta\nu\). Both magnitude and phase data are fitted with this model to determine the coupling rates accurately \cite{khalil2012analysis}. Using Eq.~(2) of the main text at a sufficiently weak probe power ($K^{0}\vert{}a_{i}\vert{}^2 \ll \kappa$), we simplify the model to a case where the first resonator exchanges energy exclusively with the transmission line and the second resonator, while the second resonator dissipates energy to the environment and not to the transmission line. The framework extends to single-sided coupled resonators, giving the scattering parameter below, in which \(g\) is the inter-resonator coupling rate, which sets the normal-mode splitting together with the bare detuning.

\begin{equation}\label{eq:S11coupled}
\begin{split}
S_{11}(\nu)=A e^{-i(2\pi\nu\tau+\phi)} \left[ -1 + \frac{\kappa_{e,1} \left( \frac{\kappa_2}{2} - i(\omega - \omega_2) \right)}{g^2 - \left( i\frac{\kappa_1}{2} + \omega - \omega_1 \right) \left( i\frac{\kappa_2}{2} + \omega - \omega_2 \right)}\right]
\end{split}
\end{equation}

\subsection*{Fitting procedure, constraints and uncertainties}
Equation~(\ref{eq:S11coupled}) is fitted to the complex \(S_{11}\) trace by nonlinear least squares over the full span containing both modes, with the free parameters \(\{\nu_1,\nu_2,g,Q_e,Q_{i1},Q_{i2},A,\phi,\tau,\delta\nu\}\). The environment factor \(\{A,\phi,\tau\}\) is first estimated from the off-resonant baseline and then released; \(\tau\) is constrained to the independently measured cable delay within \(\pm5\%\), and all quality factors are constrained positive. Probe powers are kept at least 20~dB below the onset of the Kerr-induced lineshape distortion, verified by confirming that the extracted \(\nu_0\) is power independent. Reported uncertainties are the \(1\sigma\) confidence intervals of the fit; for the rates quoted in the main text these are below \(5\%\), which is smaller than the chip-to-chip fabrication spread. All coupling and loss rates quoted in the main text for both devices are obtained in this way, from measured data --- none are taken from electromagnetic simulation. Sonnet simulations are used only to convert the measured resonance frequencies into sheet inductances (Sec.~IV).

Equation~(\ref{eq:S11coupled}) as written contains a single external coupling \(Q_e\), appropriate to the limit \(\kappa_{e,1}\gg\kappa_{e,2}\) used in the main text. Because we quote a nonzero \(\kappa_{e,2}\), the fits are performed with the second coupling retained, i.e.\ with \(\nu_2/2Q_{e2}\) added to the inner denominator of Eq.~(\ref{eq:S11coupled}); this term is small but is what makes \(\kappa_{e,2}\) an identifiable parameter, and we state it here so that the reported rates can be reproduced.

Applying this procedure to the NbTiN device gives bare frequencies \(\nu_1\) and \(\nu_2\) and coupling \(g\) that, together with the normal-mode relation given in the main text, reproduce the measured normal-mode frequencies (10.178 and 10.577~GHz) and the asymmetric linewidths. The three quantities are constrained by the measurement through \(4g^{2} = (\nu_+-\nu_-)^{2}-(\nu_1-\nu_2)^{2}\) and, independently, through the linewidth-asymmetry relation of Eq.~(7) of the main text. The unique set consistent with NbTiN (NbN) is \(\nu_1 = 10.27~\mathrm{GHz}\) (\(\nu_1 = 7.78~\mathrm{GHz}\)), \(\nu_2 = 10.49~\mathrm{GHz}\) (\(\corr{\nu_2} = 7.83~\mathrm{GHz}\)) and \(g/2\pi = 166~\mathrm{MHz}\) (\(g/2\pi = 112~\mathrm{MHz}\)), where resonator 1 --- the one nearer the feedline, with the larger \(\kappa_e\) --- is the lower in frequency; this ordering is required for the lower hybridized mode \(\nu_-\) to be the broader of the two, as observed in Fig.~2(a) of the main text \corr{(Fig.~S4(a) for NbN)}. Substituting these values into Eq.~(7) gives \(\kappa_-/2\pi = \corr{2.06}~\mathrm{MHz}\) (\(\kappa_-/2\pi = 3.87~\mathrm{MHz}\)) and \(\kappa_+/2\pi = \corr{1.45}~\mathrm{MHz}\) (\(\kappa_+/2\pi = 2.3~\mathrm{MHz}\)), consistent with the sum rule \(\kappa_++\kappa_-=\kappa_1+\kappa_2 = 2\pi\times\corr{3.51}~\mathrm{MHz}\) (\(2\pi\times 6.17~\mathrm{MHz}\)). We note for completeness that the normal-mode splitting of \(399~\mathrm{MHz}\) (\(263~\mathrm{MHz}\)) exceeds \(2g/2\pi = 333~\mathrm{MHz}\) (\(2g/2\pi = 198~\mathrm{MHz}\)): the two coincide only at the anticrossing \(\nu_1=\nu_2\). We note that the bare resonators are detuned by \(\sim220~\mathrm{MHz}\) (\(\sim150~\mathrm{MHz}\)). This detuning was not a design target: the single-sided feedline loads the nearer resonator with additional capacitance, and fabrication spread contributes the remainder. It is harmless for amplifier operation, since the 4WM condition only requires the pump to be placed at the midpoint of the hybridized modes, but it does mean that the two devices are not compared at identical hybridization angles, which we take into account when interpreting the NbN--NbTiN differences.
\\
\section{Enhancement of the Nonlinear Response}
\label{AppB}
The total inductance of a superconducting nanowire contains both a geometric (magnetic)
contribution and a kinetic contribution, \(L_T = L_{\mathrm{M}} + L_{\mathrm{K}}\).
The geometric component is set by the physical layout of the conductor and reflects the
energy stored in the magnetic field generated by the current. The kinetic inductance,
by contrast, originates from
\begin{figure}[b]
\includegraphics[width=16cm]{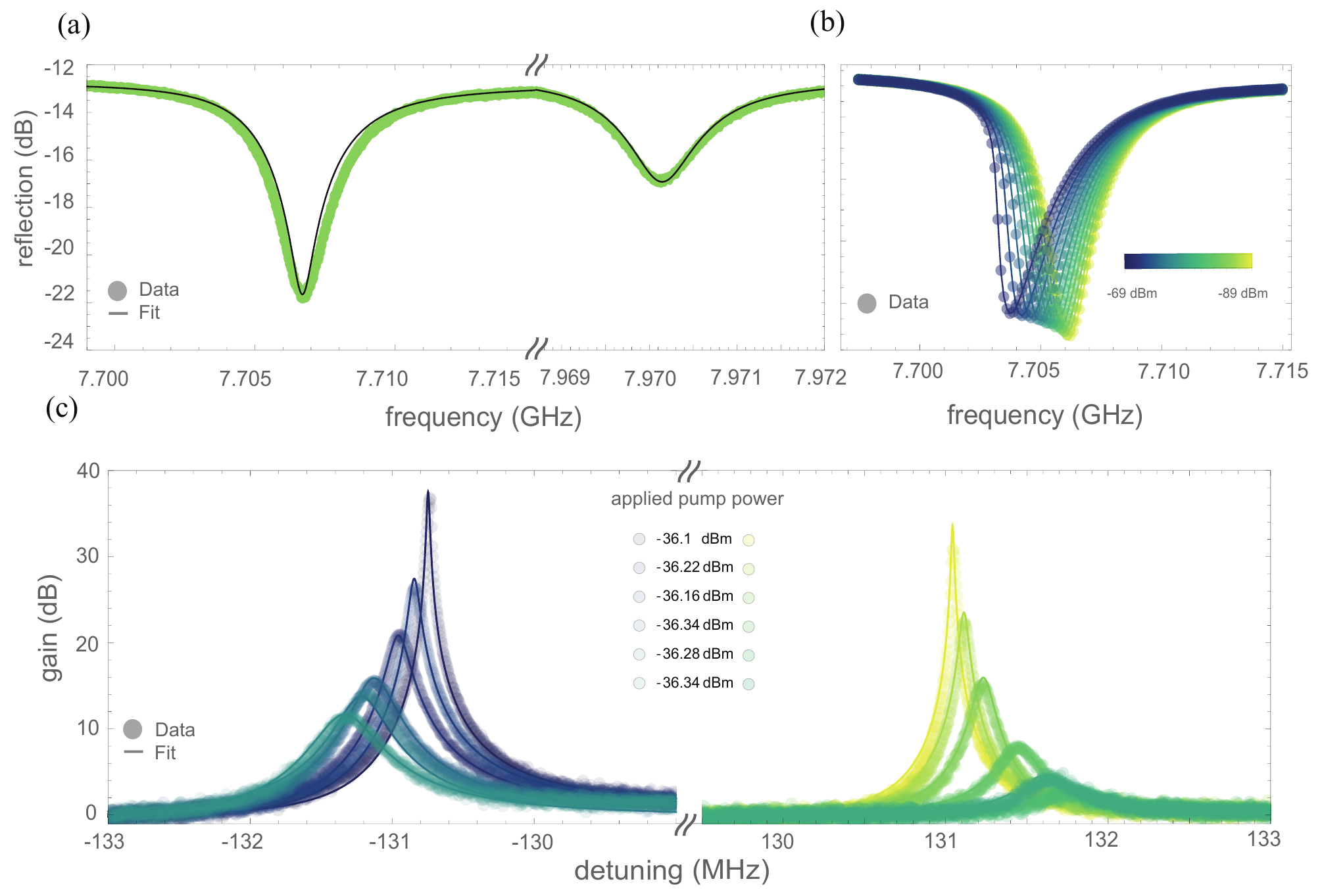}
\caption{(a) Linear microwave spectroscopy measured in reflection for the coupled NbN resonators. Solid lines are simultaneous fits of the measured complex response to Eq.~(\ref{eq:S11coupled}), from which the resonance frequencies (\(\omega_-,\omega_+\)), external coupling rates (\(\kappa_{e,1},\kappa_{e,2}\)) and internal losses (\(\kappa_{i,1},\kappa_{i,2}\)) are extracted. The fitting procedure, constraints and uncertainties are given in Sec.~III.
(b) Kerr shift of the \(\omega_{-}\) mode obtained by driving the resonator with a strong signal tone; the intracavity photon number is obtained from Eq.~(S2) using the calibrated input-line attenuation, giving \(K_0/2\pi = 0.21\)~Hz.
(c) Two-mode amplification with the pump placed midway between the hybridized modes. Points are data; solid curves are Eq.~(6) of the main text evaluated with the parameters extracted in panel (a), with only the pump amplitude adjusted. ALT TEXT: Spectral line graphs and a scatter plot demonstrating the linear microwave spectroscopy and two-mode amplification of the niobium-nitride device. The charts illustrate resonance reflection spectra, Kerr-induced frequency shifts, and pump-dependent parametric gain profiles. 
\justifying}
\label{figure33}
\end{figure}
the inertia of Cooper pairs and becomes significant in
thin, high-resistivity superconducting films where the relaxation time is long.
Enhancing this kinetic contribution is key to increasing the nonlinear response of
the resonator and reducing the pump power required to reach the amplification regime.

For a nanowire with length \(l\), width \(w\), and thickness \(t\), the kinetic inductance
can be obtained by equating the kinetic energy of the Cooper pairs to the inductive energy.
Within Bardeen–Cooper–Schrieffer theory, and for films thinner than the penetration depth,
the kinetic inductance takes the form
\[
L_{\mathrm{K}}
=
\frac{\hbar\, \rho\, l}{\pi \Delta_{0}\, w t},
\]
where \(\rho\) is the normal-state resistivity and \(\Delta_{0}\) is the superconducting
energy gap \cite{Tinkham1996}. This expression shows that narrower, thinner, and longer nanowires yield
larger kinetic inductance and therefore stronger nonlinear response.

The Kerr coefficient quantifies the resonance frequency shift per intracavity photon.
Using the quantized nonlinear resonator model, the coefficient can be expressed as
\cite{joshi2022strong}

\begin{equation}
\begin{split}
K=\frac{3}{4}\frac{\hbar\omega^2}{N_0\Delta_0^2}\frac{{(\frac{L_K}{L_T})^{2}}}{V}
\\
K=\frac{3 \hbar^{3} \rho_{n}^{\,2}}{4 N_{0} \Delta_{0}^{4} \pi^{2}}
\frac{l\,\omega^2}{(wt)^{3}\, L_T^{2} },
\end{split}
\end{equation}
The critical current does not appear explicitly here because it has been absorbed into the material parameters. The nonlinear scale current \(I^{*}\) of the current-dependent kinetic inductance \(L_k(I)=L_0[1+(I/I^{*})^{2}]\) introduced in the main text is set by the depairing current of the film: within Ginzburg--Landau theory the two differ only by a numerical factor of order unity that is independent of the material parameters \cite{Annunziata_2010,zmuidzinas2012superconducting}. Substituting \(I^{*}\propto I_c\) into \(L_k(I)=L_0[1+(I/I^{*})^{2}]\) and quantizing reproduces the expression above with \(K\propto \omega^{2}(L_K/L_T)^{2}/(L_K I_c^{2})\), so the form written in terms of \(\rho_n\), \(\Delta_0\) and the wire dimensions is equivalent to one containing \(I_c\) explicitly, not an approximation that neglects it. The same relation is what allows the intracavity current at compression to be quoted as a fraction of \(I^{*}\simeq I_c\) in Sec.~V without an independent measurement of \(I_c\).
Here \(N_{0}\) is the density of states at the Fermi level, \(\omega\) is the resonant
frequency, and \(L_T\) is the total inductance. This expression makes clear that the Kerr
coefficient increases with kinetic-inductance participation, higher resonant frequency,
smaller cross-sectional area, and smaller superconducting energy gap. We extracted sheet inductances \({L}^k_\Box\)= 103 pH/$\Box$ and \({L}^k_\Box\)= 31 pH/$\Box$ for ten-nanometre-thick NbN and NbTiN films, respectively, by fitting the measured resonance frequencies of our designed single and coupled resonators using Sonnet electromagnetic (EM) simulations.
Motivated by these scalings, we implemented the amplifier design in niobium–nitride, a
material with a smaller superconducting energy gap than niobium–titanium–nitride
\cite{tripathy2024comprehensive}. This choice increases the kinetic inductance and
strengthens the nonlinear response, allowing parametric amplification to be achieved
with substantially lower pump power.

\section{Comparing the NbN and NbTiN devices}
Although drawn from the same mask, the two devices are not equivalent circuits. Because the sheet inductance differs by a factor \(103/31 = 3.3\), the same layout produces different total inductance, participation ratio, characteristic impedance and resonance frequency (7.84~GHz for NbN versus 10.38~GHz for NbTiN). We therefore treat the comparison as one between two devices of the same design, and make the parameter dependence explicit for each figure of merit rather than attributing differences to the film alone.

\emph{Participation ratios.} The comparison can be made quantitative without any additional measurement, because the two devices share a mask and therefore share their geometric inductance \(L_M\) and their capacitance \(C\). The measured resonance frequencies then fix the total inductances: \(L_T^{\mathrm{NbN}}/L_T^{\mathrm{NbTiN}} = (\bar\omega_{\mathrm{NbTiN}}/\bar\omega_{\mathrm{NbN}})^{2} = (10.38/7.84)^{2} = 1.75\). Combining this with the independently extracted sheet inductances, \(L_K^{\mathrm{NbN}}/L_K^{\mathrm{NbTiN}} = 103/31 = 3.32\), and with \(L_T = L_M + L_K\) gives a single solution,
\begin{equation}
\frac{L_K}{L_T}\bigg|_{\mathrm{NbTiN}} = 0.32, \qquad
\frac{L_K}{L_T}\bigg|_{\mathrm{NbN}} = 0.61 .
\end{equation}
These are the values listed in Table~\ref{tab:comparison}. Note in particular that the participation ratio does not scale with the sheet inductance: it rises by a factor \(1.9\), not \(3.3\), because a substantial geometric inductance remains in series. Any comparison that assumes \(L_K/L_T \simeq 1\) for both films, or that propagates the raw sheet-inductance ratio into a participation-ratio scaling, will be wrong by an order of magnitude in \(K\).

\emph{Kerr coefficient.} With \(K \propto \omega^{2}(L_K/L_T)^{2}/V\) at fixed material parameters and fixed mode volume \(V=lwt\), the circuit differences alone predict a ratio
\begin{equation}
\frac{K_{\mathrm{NbN}}}{K_{\mathrm{NbTiN}}}\bigg|_{\mathrm{circuit}}
= \left(\frac{7.84}{10.38}\right)^{2}\left(\frac{0.61}{0.32}\right)^{2} = 2.0 ,
\end{equation}
against a measured ratio of \(0.21/0.01 = 21\). Circuit parameters therefore account for only a factor \(2\) of the observed twenty-fold difference. The remaining factor of \(\sim10\) must come from the material prefactor of Eq.~(S5). It is worth being explicit about what is and is not already counted here: the normal-state resistivity and the gap enter the sheet kinetic inductance, \(L^{k}_\Box=\hbar\rho_n/(\pi\Delta_0 t)\), and hence are already contained in the participation ratio used above, so they must not be counted twice. With \(\omega\), \(L_K/L_T\) and the wire cross-section all fixed by measurement, the only residual material dependence in Eq.~(S5) is \(K\propto 1/(N_0\Delta_0^{2})\); a factor \(\sim10\) then requires \(N_0\Delta_0^{2}\) to be roughly an order of magnitude smaller in our NbN film than in our NbTiN film. The smaller superconducting gap of NbN acts in this direction, as would a chip-to-chip difference in the \(\sim20\%\) nanowire-width bias of Fig.~\ref{figSEM}(c), which enters both through the mode volume and through the participation ratio, but we cannot verify the size of the effect with the present data. We deliberately do not tune these material parameters to close the gap: the honest conclusion is that the enhancement is predominantly a material effect rather than a circuit effect, but that our data do not determine \(\Delta_0\), \(\rho_n\) and \(N_0\) for these particular films well enough to verify the prefactor quantitatively. Doing so would require independent transport and tunnelling characterization of the two films, which we did not perform.

\emph{Saturation power.} Gain compression sets in when the Kerr-induced detuning approaches the linewidth, \(K n \sim \kappa\). Using \(n = 4\kappa_e P_{\mathrm{in}}/(\hbar\omega\kappa^{2})\) gives
\begin{equation}
P_{\mathrm{1dB}} \;\simeq\; \frac{\hbar\omega\,\kappa^{3}}{4\,\kappa_{e}\,K}\,\eta(G),
\end{equation}
with \(\eta(G)\) a gain-dependent factor of order unity. Saturation is therefore not determined by \(K\) alone: a device with a larger Kerr coefficient can still saturate at higher power if its linewidth is correspondingly larger. Evaluating the ratio term by term with the measured parameters (\(\kappa/2\pi = 3.4\) and \(2.31~\mathrm{MHz}\), \(\kappa_e/2\pi = 2.3\) and \(1.40~\mathrm{MHz}\), \(K/2\pi = 0.21\) and \(0.01~\mathrm{Hz}\) for NbN and NbTiN respectively),
\begin{equation}
\frac{P_{\mathrm{1dB}}^{\mathrm{NbN}}}{P_{\mathrm{1dB}}^{\mathrm{NbTiN}}}
= \underbrace{0.76}_{\hbar\omega}\times
\underbrace{1.94}_{\kappa^{3}/\kappa_e}\times
\underbrace{0.048}_{1/K}
= 0.070 \;\;(-11.6~\mathrm{dB}).
\end{equation}
The measured difference is \(\corr{-91-(-81.5) = -9.5}~\mathrm{dB}\). The scaling thus reproduces the sign and the order of magnitude; the residual \(\corr{2}~\mathrm{dB}\) is accounted for by the \(\pm1.5~\mathrm{dB}\) input-line calibration on each device and by \(\eta(G)\), which depends on the hybridization angle and therefore differs between the two devices. The essential point is the ordering: the device with the larger Kerr coefficient saturates at the lower power, as expected, and the saturation values listed in Table~\ref{tab:comparison} are consistent with this scaling.

\emph{Critical current.} The ratio of the intracavity current to the depairing current at compression can be obtained from the same measured quantities, without an independent determination of \(I_c\). Writing the current-induced pull of the resonance as \(\omega(I) \simeq \omega_0[1-(p/2)(I/I^{*})^{2}]\) with \(p = L_K/L_T\), the compression condition \(K n \sim \kappa\) is equivalent to a Kerr-induced detuning of one linewidth, so
\begin{equation}
\left(\frac{I}{I^{*}}\right)^{2} = \frac{2\kappa}{p\,\omega_0} .
\end{equation}
With \(p = 0.32\), \(\kappa/2\pi = 2.31~\mathrm{MHz}\) and \(\omega_0/2\pi = 10.38~\mathrm{GHz}\) this gives \(I/I^{*} = 0.037\) for NbTiN, and with \(p = 0.61\), \(\kappa/2\pi = 3.4~\mathrm{MHz}\), \(\omega_0/2\pi = 7.84~\mathrm{GHz}\) it gives \(I/I^{*} = 0.038\) for NbN. The near-coincidence of the two values is not required by the physics --- the combination \(2\kappa/(p\omega_0)\) simply happens to be almost equal for the two devices --- but it does show that both compress at a comparable fraction of the depairing current, each when its own Kerr shift reaches its own linewidth. Since \(I^{*}\) and \(I_c\) differ only by a material-independent numerical factor of order unity, the intracavity current at the compression point is below \(10\%\) of the depairing current in both devices. Saturation is thus governed by the Kerr-induced frequency pull of the resonant mode and not by the film critical current, which is why increasing \(I_c\) alone would not raise \(P_{\mathrm{1dB}}\), whereas increasing \(\kappa\) at fixed \(K\) would. For orientation, the \(1~\mu\mathrm{m}\times10~\mathrm{nm}\) nanowire cross-section and depairing current densities of \(10^{6}\)--\(10^{7}~\mathrm{A\,cm^{-2}}\) typical of these films correspond to \(I_c \sim 0.1\)--\(1~\mathrm{mA}\), so the intracavity current at compression is of order a few to a few tens of microamperes. 

\emph{External coupling.} For a lumped resonator coupled through \(C_c\) to a line of impedance \(Z_0\), \(\kappa_e = \omega^{2}C_c^{2}Z_0/C_\Sigma\). Since \(C_c\) and \(C_\Sigma\) are geometric and therefore common to the two chips, an identical layout predicts \(\kappa_e \propto \omega^{2}\), i.e.\ \(\kappa_e^{\mathrm{NbN}}/\kappa_e^{\mathrm{NbTiN}} = 0.57\). The measured ratio is \(2.3/1.40 = 1.6\), a discrepancy of \(2.9\), corresponding to an effective \(C_c\) larger by \(\sqrt{2.9} = 1.7\) in the NbN chip. We attribute this to the lithographic width bias documented in Sec.~I --- metal features are systematically wider than drawn, which narrows the capacitor gaps and raises \(C_c\) --- rather than to the film. This is the clearest single illustration of why the two devices cannot be treated as a controlled material comparison: a figure of merit that depends on the fabricated gap dimension varies chip to chip by more than the frequency-scaling effect one would attribute to the material.
\\
\emph{Consequences for the material comparison.} The three figures of merit therefore behave differently. The Kerr coefficient is dominated by material properties: circuit parameters account for a factor \(2\) out of the measured \(21\). The external coupling and, through it, the gain--bandwidth product and the saturation power are dominated by circuit and fabrication parameters, and their differences should not be read as material properties at all. Accordingly we make no claim in the main text that the observed differences isolate an intrinsic property of NbN versus NbTiN. A controlled comparison would require either films of matched sheet inductance, or geometries adjusted so that the two amplifiers operate at the same frequency with the same participation ratio, together with independent transport characterization of \(\rho_n\), \(T_c\) and \(I_c\) for each film. We regard this as the natural next step and state it as such rather than implying it has been done here.
\\
\section{Required conditions for amplification in NbTiN and NbN }
Operating a KIPA in the amplification regime requires driving the device at a precise pump frequency and power. Figure~\ref{figure44}(a) and (b) show the pump-power sweep for the highest gain of both modes: at a particular power, enough field is confined in the cavity to generate photon pairs. In panels (c) and (d) the drive frequency is tuned; the dotted line marks the optimum at 10.3701~GHz, \(7.4~\mathrm{MHz}\) below the midpoint of the two hybridized modes (\(10.3775~\mathrm{GHz}\)), \corr{which we attribute to} the current-induced downward pull of both modes at the operating pump power. Increasing the pump power further drives the coupled nonlinear resonators into a bistable regime with strong field confinement in both cavities.

In Fig.~\ref{figure33}, the NbN device was characterized using the same methods described for the
NbTiN sample. Reflection measurements show two hybridized modes at \(7.707~\mathrm{GHz}\) and \(7.970~\mathrm{GHz}\), from which the
coupling rates were extracted using the complex-\(S_{11}\) fit of Sec.~III. Power-dependent measurements of the
lower mode confirmed the expected Kerr-induced frequency shift used to extract the self-Kerr coefficient. Driving the circuit with
a pump tone at the mean frequency of the hybridized modes generates nondegenerate
signal and idler peaks. Sweeping the pump power from \(-36.34~\mathrm{dBm}\) to
\(-36.10~\mathrm{dBm}\) enhanced the parametric gain, reaching nearly 40 dB, as
shown in Fig.~\ref{figure33}(c). Table~\ref{tab:comparison} compares NbN and NbTiN with other similar devices. In both devices the calibrated pump power and frequency were tuned to maximize the peak parametric gain, in order to demonstrate the upper limit achievable with these films in this architecture. We chose peak gain as the optimization target because the purpose of these measurements is to establish that limit, not to define a practical operating point. For multiplexed readout the relevant target is instead the gain--bandwidth product at a required gain (typically 20~dB), which is maximized at lower pump power and a slightly detuned pump: Fig.~3(b) of the main text shows that the GBP is essentially flat over 15--40~dB of gain for both devices, so operating at 20~dB rather than at peak gain costs little bandwidth while leaving substantially more headroom before compression. Optimizing the pump for GBP rather than for peak gain therefore changes the operating point but not the reported figures of merit.

\section{Uncertainty budget}
\label{sec:uncertainty}
The table below sets out the full uncertainty budget for every quantity quoted in the main text and in Table~\ref{tab:comparison}.

\begin{table*}[h]
\caption{Uncertainty budget for the quantities quoted in the main text and in Table~\ref{tab:comparison}. ALT TEXT: A data table outlining the complete uncertainty budget for the measured and calculated parameters of the amplifier devices. It details the dominant statistical and systematic errors contributing to the net confidence intervals.}
\begin{ruledtabular}
\begin{tabular}{lllc}
\textrm{Quantity} & \textrm{Dominant statistical term} & \textrm{Dominant systematic term} & \textrm{Net \(1\sigma\)} \\
\colrule
Resonance frequencies \(\omega_\pm\) & fit s.e., \(\kappa/2\mathrm{SNR}\) & VNA frequency reference, negligible & \(<10^{-4}\) \\
Coupling rates \(\kappa_{e,j},\kappa_{i,j}\) & \(1\sigma\) fit confidence interval & choice of baseline window & \(<5\%\) \\
Exchange coupling \(g\) & propagated from \(\omega_\pm,\omega_{1,2}\) & --- & \(<5\%\) \\
Intracavity photon number \(n\) & SA power reading, \(<0.1\)~dB & input-line calibration, \(\pm1.5\)~dB & \(\times\!\div1.4\) \\
Kerr coefficient \(K\) & slope fit, \(<3\%\) & inherited from \(n\) & \(\times\!\div1.4\) \\
Gain \(G\) & pump ON/OFF ratio, \(<0.1\)~dB & cancels in the ratio & \(\pm0.2\)~dB \\
Bandwidth \(\Gamma\) & Lorentzian FWHM s.e. & --- & \(\lesssim4\%\) \\
Gain--bandwidth product & propagated from \(G,\Gamma\) & --- & \(\lesssim5\%\) \\
\(P_{\mathrm{1dB}}\) & gain-curve interpolation, \(\pm0.5\)~dB & input-line calibration, \(\pm1.5\)~dB & \(\pm1.6\)~dB \\
Output-chain gain \(\corr{G}\) & \(1\sigma\) fit confidence interval & VTS thermometry, \(\pm5\)~mK & \(\pm0.3\)~dB \\
System added noise \(\corr{n_{\mathrm{add}}}\) & \(1\sigma\) fit confidence interval & VTS thermometry & \(\pm0.4\) quanta \\
Sheet inductance \(L_\Box^{k}\) & propagated from \(\omega_0\) & Sonnet geometry, incl.\ etch bias & \(\pm8\%\) \\
Participation ratio \(L_K/L_T\) & propagated from \(\omega_0\), \(L_\Box^{k}\) & assumption of common \(L_M\), \(C\) & \(\pm0.05\) \\
\end{tabular}
\end{ruledtabular}
\label{tab:uncert}
\end{table*}

Three features of this budget deserve emphasis:
First, in Fig.~2(b) the abscissa is the intracavity photon number, which is obtained from the calibrated input-line attenuation. Its \(\pm1.5~\mathrm{dB}\) uncertainty is a single, common scaling of every point on the axis. It does not scatter the points relative to one another --- the linearity of the Kerr shift, which is the physical content of the figure, is unaffected --- but it tilts the fitted slope by \(\times\!\div1.4\). We therefore quote \(K/2\pi = 0.010^{+0.004}_{-0.003}~\mathrm{Hz}\) (NbTiN) and \(0.21^{+0.09}_{-0.06}~\mathrm{Hz}\) (NbN) rather than drawing per-point bars that would be both invisible and misleading. We note that this systematic is common to the two devices, so it cancels in the \(K\) ratio used in Sec.~V, which is accurate to the statistical level of a few percent.
Second, in Fig.~3(b) each point is a Lorentzian FWHM extracted from an independently acquired gain profile at a fixed pump power. The \(1\sigma\) standard error of that fit is \(\lesssim4\%\), which combined with the \(\pm0.2~\mathrm{dB}\) gain calibration gives \(\lesssim5\%\) on the gain--bandwidth product --- comparable to the marker size. The relevant systematic, the input-line calibration, shifts the pump power at which a given gain is reached but not the gain or the bandwidth measured there, and so does not enter the plotted quantity at all.
Third, in Fig.~\ref{figsetup}(b) the ordinate is an average of \(500\) spectra in a \(100~\mathrm{Hz}\) bandwidth, giving a standard error below \(1\%\), and the abscissa is a thermometer reading with \(\pm5~\mathrm{mK}\) uncertainty, negligible over the \(50\)--\(1700~\mathrm{mK}\) range plotted. The quantities of interest are not the individual points but the two fitted parameters, \corr{whose confidence intervals are quoted in the caption of Fig.~S2}.

\bibliography{References}